\documentclass[usenatbib]{mnras}

\usepackage[T1]{fontenc}

\usepackage{graphicx}	
\usepackage{amsmath}	
\usepackage{amssymb}	
\usepackage{verbatim} 
\usepackage{lscape}
\usepackage{dcolumn}
\usepackage{color}
\usepackage{float}
\usepackage{subfig}
\usepackage{tabularx}
\usepackage{soul}
\usepackage{ulem}
\definecolor{jm}{rgb}{0.23, 0.53, 0.75}

\title[Power spectrum of density fluctuations with primordial black holes]{Power spectrum of density fluctuations, halo abundances and clustering with primordial black holes}

\author[N. D. Padilla et al.]{
Nelson D. Padilla,$^{1,2}$\thanks{E-mail: npadilla@astro.puc.cl}
Juan Maga\~na,$^{1,2}$
Joaqu\'\i n Sureda$^{1,2}$
and Ignacio J. Araya$^{3}$
\\
$^{1}$Instituto de Astrof\'\i sica, Pontificia Universidad Cat\'olica de Chile, Vicu\~na Mackenna 4860, Santiago, Chile\\
$^{2}$Centro de Astro-Ingenier\'\i a, Pontificia Universidad Cat\'olica de Chile, Vicu\~na Mackenna 4860, Santiago, Chile\\
$^{3}$Instituto de Ciencias Exactas y Naturales, Facultad de Ciencias, Universidad Arturo Prat,\\ Avenida Arturo Prat Chac\'on 2120, 1110939, Iquique, Chile
}

\date{Accepted 2021 April 16. Received 2021 April 13; in original form 2020 December 16}

\pubyear{2021}

\begin{document}
\label{firstpage}
\pagerange{\pageref{firstpage}--\pageref{lastpage}}
\maketitle

\begin{abstract}
We calculate the effect of dark matter (DM) being encapsulated in primordial black holes (PBHs) on the power spectrum of density fluctuations $P(k)$; we also look at its effect on the abundance of haloes and their clustering. We adopt a growth of Poisson fluctuations that starts only after the moment of matter and radiation equality and study both monochromatic and extended Press-Schechter distributions. We present updated monochromatic black hole mass constraints by demanding $<10\%$ deviations from the $\Lambda$ cold dark matter power spectrum at a scale of $k=1$hMpc$^{-1}$.  Our results show that PBHs with masses $>10^4$h$^{-1}M_\odot$ are excluded from conforming all of the dark matter in the Universe; they would, however, be able to act as seeds for super massive black holes if they conform a small fraction of the DM.  We also apply this condition to our extended Press-Schechter (PS) mass functions, and find that the Poisson power is scale dependent even before applying evolution, due to the change of the mass density in PBHs with redshift, and therefore with scale, as they start affecting the gravitational potential at different times. We find that characteristic masses $M^*\leq10^2 $h$^{-1}M_\odot$ are allowed, {leaving only two characteristic PBH mass windows of PS mass functions when combining with previous constraints, at $M^*\sim10^2$h$^{-1}M_\odot$ and $\sim10^{-8}$h$^{-1}M_\odot$ where all of the DM can be in PBHs.  The resulting dark matter halo mass functions within these windows are similar} to those resulting from cold dark matter made of fundamental particles, but as soon as the parameters produce unrealistic $P(k)$, the resulting halo mass functions and their bias as a function of halo mass deviate strongly from the behaviour measured in the real Universe.
\end{abstract}


\section{Introduction}

Our current cosmological paradigm posits a flat Universe that is currently dominated by dark energy, with about one third of the energy density in the form of matter, of which $\sim 80\%$ is made up of an unknown form which does not interact with photons called dark matter, and the remaining $\sim 20\%$ corresponding to ordinary baryons.  The accuracy with which this model fits the power spectrum of temperature fluctuations of the cosmic microwave background, the CMB \citep{PlanckX2018}, is stunning and indicates that at the very least it represents a very effective model of the Universe \citep{Peebles2020}.  

Because of this there has been a lot of effort trying to understand the nature of these two dark components.  Here we will concentrate on an alternative scenario for the dark matter.  Even though from the particle physics point of view there are excellent candidates for the dark matter particle in Supersymmetry (see \citealt{deSwart2017} for a historical review), it is not clear whether dark matter would be made up of a single particle (P.J.E. Peebles, private communication).  
\vskip -.05cm

In line with this possibility, there has been a lot of attention  devoted  to primordial black holes (PBHs).  These objects are thought to be able to form early in the history of the Universe as noted early on by \cite{Zeldovich1966}, \cite{Hawking1971} and \cite{Carr_Hawking:1974}.  In these works the authors noticed the possibility that overdensities in the early universe could collapse and form black holes.  Since then there have been numerous works that have explored the possibility that these PBHs could make up a large fraction of the dark matter, even perhaps all of it (see for instance \citealt{Khlopov_2010,Carr:2016,Carr:2017,Carr:2020}). In order to test this idea the MAssive Compact Halo Object searches (MACHO), proposed initially by \cite{Paczynski1986} were designed and executed in the nineties.  The result of the MACHO searches returned such low occurrence of detected microlensing events that it led to the idea that PBHs could not possibly make up a sizeable fraction of the dark matter (see for instance \citealt{Alcock2000,Ogle:2011,Tisserand:2007}).  However, after the first detection of a gravitational wave event by the Laser Interferometer Gravitational-Wave Observatory (LIGO) consortium \citep{Abbott:2016blz}, that included black holes with masses that could not be comfortably produced from the death of massive stars in standard stellar evolution scenarios, the interest in PBHs  rekindled rapidly (see for instance \citealt{Garcia-Bellido:2017fdg,Bird2016}).

As PBHs do not form from dying stars, they can have any range of masses.  There is a simple way to estimate the minimum mass a PBH { can have by requiring that its Schwarzschild radius be equal to its Compton wavelength.  This leads to  $M_{\rm PBH}\sim 10^{-5}$g, i.e. one Planck mass}.  In the case where the PBH forms as a fluctuation collapses in the early Universe, the maximum mass can be even larger than the mass within the present-day horizon, even though these objects would naturally lie outside our own horizon. In most of the literature about PBHs it is assumed that they  evaporate due to Hawking radiation \citep{Hawking:1974Nature,Hawking:1975}, i.e., the mass of a PBH is not constant with time, and the mass of a PBH determines the time it will survive since its formation before it evaporates away.  A PBH of $\sim 1.7 \times 10^{11}$g formed at the end of inflation would be in their last stages of evaporation by the present day.

The PBH mass spectrum depends on the mechanism that leads to their formation.  In the case where there is a characteristic, single scale for the collapse of PBHs, their masses would be unique, in what is commonly termed monochromatic mass functions \citep{Dolgov:1993PhRvD, Garcia-Bellido:1996PhRvD, Clesse:2015PhRvD, Green:2016PhRvD, Inomata:2017PhRvD}.  This characteristic scale arises as a result of the need to increase the fluctuations in the energy content of the Universe in these early epochs, because the physical overdensity needed for collapse is actually quite large of $\delta_c^{\rm physical}\sim 0.5$ \citep{Green:1999PhRvD,Niemeyer:1998PhRvL}.  However, it is clear that these ideas are simplified and the actual distribution of PBH masses can be broad, to different degrees.  The actual abundance of PBHs can be obtained by applying peaks theory approaches to the collapse of these overdensities \citep{Germani:2019,Germani:2020,Escriva:2020}.

The difficulty in reaching the large fluctuations required for black hole collapse as the overdense patch enters the horizon can  be alleviated by an increase in the small scale power in the primordial spectrum.  For instance, hybrid inflation models produce what is usually termed a blue index \citep{Linde:1994PhRvD}, and this has been the subject of several studies for formation of PBHs (see for example \citealt{Kawasaki:2013PhRvD,Gupta:2018}).

In a recent paper by \cite{Sureda:2020} (SMAP from this point forwards) we presented a study of extended distributions of mass of PBHs formed from primordial inhomogeneities with such blue indices in two different scenarios.  One of them corresponds to the formation of PBHs in 
bubble collisions during inflation \citep{Hawking:1982_bubble,Kodama:1982sf,Deng:2017uwc, Deng:2020arXiv} which includes a novel consideration that the collisions of bubbles will occur in the anti-peaks of the density field at these early times.  Since the positive and negavite peaks of a linear density field distribute equally, in SMAP we take the primordial power spectrum with a blue tilt as the source of these fluctuations and apply the Press-Schechter (PS) formalism \citep{press1974} to obtain a  PBH mass function for what is termed the fixed conformal time formation mechanism (FCT).  In SMAP we also take the route of using PS to obtain the PBH mass function for formation at horizon crossing (HC).  

With the possibility, at least from the theoretical point of view, of producing these primordial black holes, also came the study of the consequence of their existence.  Assuming that PBHs emit radiation due to their evaporation, it is possible to estimate how much energy would be injected to the medium due to this process, in particular during Big-Bang Nucleosynthesis, which could affect the abundance of light elements \citep{Carr:2016} or as an excess of extragalactic
$\gamma$-rays \citep{Carr:2016_gammaray}.  The gravitational effects of PBHs can be seen via lensing of background objects \citep{Hawkins:2020arXiv200107633H}, or because of the effects of their accretion by neutron stars \citep{Capela:2013}, among several other processes.  There are also constraints related to gravitational waves, for example so as not to exceed the detected events from LIGO \citep{Abbott:2016blz} presented by \cite{Ali-Haimoud:2017}, from the lack of detections of lower mass black hole mergers by \cite{Abbott:2018}, and from the gravitational wave background limits resulting from the first Advanced LIGO run \citep{Wang:2018}.  
Notice that some of these constraints are still under study as \cite{Montero-Camacho:2019} show that the neutron stars constraints should be relaxed, which is also the case with the constraints from gravitational wave detections \citep{Boehm:2020}.

Constraints on the amplitude of the power spectrum on small scales also restrict the possibility that PBHs can form from large non-linear anisotropies in the power spectrum (e.g. \citealt{Gow2020}). The effect of large overdensities of massive PBHs induce
$\mu$ distortions in the CMB \citep{Chluba_2015}.
On the other hand, the production of primordial gravitational waves from large amplitude fluctuations cannot exceed Pulsar Timing Array (PTA) thresholds, updated with the North American Nanohertz Observatory for Gravitational waves (NANOGrav,
\citealt{Arzoumanian2018}), and fluctuations cannot exceed the
small-scale constraints on the power spectrum from y-distortions \citep{Lucca_2020}, or $21$cm observations \citep{Gong_2017,Bernal_2018,Mena_2019,Mu_oz_2020}.  Put together these constraints place an upper limit of $100-10^4M_\odot$ for the mass of PBHs of monochromatic or narrow log-normal mass distributions centered on these masses.
Future constraints from the Primordial Inflation eXplorer (PIXIE) will put more stringent constraints on the amplitude of spikes in the spectrum, which will not be able to exceed the current measured amplitude of the CMB spectrum by more than one order of magnitude up to scales of $k\sim 10^4$hMpc$^{-1}$  \citep{Chluba2012}.
The presence of PBHs as a fraction of the DM can also induce the formation of mini-haloes which would boost the fluctuations spectrum.  The consequences of these have also been used to put similar constraints on the possible abundance of PBHs (e.g. \citealt{Gosenca_2017}).

In general, adding all these constraints together gives us an idea of the allowed range of PBH masses that can constitute a sizeable fraction of the dark matter.  These constraints can  be translated and applied to extended mass distributions (\citealt{Bellomo_2018}, \citealt{Carr:2017}, SMAP).  In particular, SMAP show that combining all available monochromatic constraints that are not currently in dispute, and including an additional one for the abundance of super massive black holes of Active Galactic Nuclei applicable specifically to extended PBH mass distributions, there are several regions of the parameter space for the FCT and HC scenarios that still allow $100\%$ of the dark matter in PBHs.

This paper explores the effects of stochasticity of having primordial black holes make up a fraction $f_{\rm PBH}$ of dark matter in the Universe.  We concentrate on effects on the power spectrum of density fluctuations $P(k)$ and place constraints on the abundance of PBHs that are complementary to these previous works, concentrating on scales accessible via large scale structure measurements.  We also study the abundance and clustering of dark matter haloes of mass $M_{h}$  (\citealt{Carr_2018,Carr:2020arXiv200212778C}), and how these constrain the fraction of mass in PBHs.  The mass distributions of PBHs come either in monochromatic or extended form, which will have different phenomenology and will be explored separately.  In the case of extended distributions we will verify whether allowed regions of the parameter space from SMAP are still viable.

This paper is organized as follows.  In Section \ref{sec:mf} we present the extended mass functions for PBHs, to then calculate the effect of Poisson noise on the power spectrum of density fluctuations in Section \ref{theory}, where we also look at the case of monochromatic distributions.  In Section \ref{dndm} we use the power spectrum including Poisson noise to infer the resulting halo mass functions and dependence of  bias on halo mass.  Section \ref{pofk} shows the resulting constraints on the fraction of dark matter that can be in the form of PBHs with monochromatic and extended mass distributions.  We finish summarising our work in Section \ref{sec:conclusions}.

\section{Extended primordial black hole mass functions}
\label{sec:mf}

We will use the extended mass functions of a previous paper of our team, SMAP, where we use a modified Press-Schechter (PS) formalism (e.g. \citealt{Sheth2001}) to obtain the mass functions for two different PBH formation mechanisms.

In the PS formalism, the differential mass function takes the form,
\begin{equation}
\frac{dn}{dM}\left(  M\right)  =A_n \nu f\left(  \nu\right)  \frac{\overline
{\rho}_m}{M^{2}}\frac{d\log\nu}{d\log M}=A_n f\left(  \nu\right)  \frac
{\bar\rho_m}{M}\frac{d\nu}{dM},
\label{eq:dndm_gen}
\end{equation}
\noindent
where $M$ is the PBH mass, $\bar\rho_m$ is the average density in PBHs, $A_n$ is the amplitude which is set such that $\bar\rho_m$ is the  matter density at $z=0$ (see below), and 
the multiplicity function can  be expressed as,
\begin{equation}
f(\nu(M))=\frac{2}{\sqrt{2\pi}}\exp \left(-\frac{1}{2}  \nu(M)^{2} \right).
\end{equation}
Notice that this function is normalized such that its integral from $0$ to $\infty$ is $1$. In these expressions, $\nu(M)$ is a measure of the peak height defined as
\begin{equation}
\nu\left(  M\right)
=\frac{\delta_{c}}{\sigma\left(  M\right) },
\label{eq:nu}
\end{equation}
where $\delta_{c}$ is the linear threshold for PBH formation, and $\sigma^2(M)$ the variance of the density field which depends on the power spectrum of density fluctuations.  

In the case of black holes that form in the early universe, the relevant power spectrum is the primordial one which consists on a power law that allows, beyond observable scales, features such as the primordial spectral index becoming larger or, equivalently, bluer \citep{Linde:1994PhRvD}, 
\begin{equation}
P_{\rm prim}(k) = \begin{cases}
A\left(\frac{k}{k_0}\right)^{ns} &\text{for $k<k_{\rm piv}$},\\
A\epsilon\, \left(\frac{k}{k_0}\right)^{n_{b}} &\text{for $k\geq k_{\rm piv}$}.
\end{cases}
\label{eq:pkbroken}
\end{equation}
The normalization is set at the fiducial wavenumber, $k_0=0.05\,\mathrm{ Mpc}^{-1}$, for which $A = 2.101\times 10^9 \,\mathrm{ Mpc}^3$, $n_{s}=0.9649\pm 0.00042$, consistent with no evidence for significant deviation from a power-law over the range $0.008 \mathrm{Mpc^{-1}} \leq k \leq 0.1\mathrm{Mpc^{-1}}$ \citep{PlanckX2018}.
The pivot in the power law is introduced at the scale $k_{\rm piv}$ above which the spectral index turns blue, i.e. $n_{b}>1$; we choose $k_{\rm piv}=10$Mpc$^{-1}$, well beyond the observable range of the linear power spectrum in CMB and large scale structure measurements (e.g. \citealt{Ivanov2019} reach $k=0.25$hMpc$^{-1}$).
Since $P(k)$ must be a continuous function 
$$A\, \left(\frac{k_{\rm piv}}{k_0}\right)^{n_s}=A \,\epsilon \left(\frac{k_{\rm piv}}{k_0}\right)^{n_{b}}$$ and
 $$\epsilon=\left(\frac{k_{\rm piv}}{k_0}\right)^{n_s-n_{b}}.$$

We assume that black holes evaporate due to Hawking radiation which translates into no black holes with masses below the mass that has evaporated at a given redshift, i.e. because of this the actual PBH mass function evolves with time as the number density at  low PBH masses will change depending on whether that mass has already evaporated (zero number density) or not.  Additionally, we only take into account in our calculations black holes with an abundance of at least one per Hubble volume.  This last consideration is not a modification to the mass function, only a realization that the mass within the horizon is the only relevant one at any given moment.  These two effects make the actual mass density of PBHs  different from $\bar\rho$ unless the extra factor $A_n$ is added to ensure consistency with the average mass density (see the following subsection for more details).

Therefore, extended mass distributions take the following generic form,
\begin{equation}
dn/dM(M,z) = \begin{cases}
0 &\text{for $M<M_{ev}(z)$},\\
F_{\rm process}(n_{b}, M^*, k_{\rm piv}) &\text{for $M\geq M_{ev}(z)$},
\end{cases}
\end{equation}
where $M_{ev}(z)$ is the PBH mass that evaporates completely by redshift $z$ and $F_{\rm process}$ is a linear combination of functions of the Schechter form fully determined by the process of PBH formation, the blue spectral index of the primordial power spectrum, $n_{b}$, which characterises the primordial spectrum above 
 the pivot wavenumber $k_{\rm piv}$, and
the characteristic mass scale, $M^*$, and can be found in \cite{Sureda:2020}.

\subsection{PBH formation at fixed conformal time and at horizon crossing}
The processes for formation presented in SMAP consist of the following.

(i) Formation of PBH as a fluctuation enters the horizon.  For this the amplitude of fluctuations in the PS formalism are calculated at the time the mode enters into causal contact.  This channel of formation is referred to as "Horizon Crossing" or HC.  In SMAP we assume that the PBH formation occurs only during the epoch of radiation domination.  Note that in no cases would a PBH form after equality if $k_{\rm piv}>k_{eq}$, where $k_{eq}\sim 2\pi/R_H(z=z_{eq})$ is the wavenumber corresponding to the horizon radius at the time of equality $z_{eq}$.  The mass within the scale corresponding to $k_{\rm piv}$ is the maximum mass that a PBH can take in this formalism, as at smaller wavenumbers the power spectrum index is $n_s=0.96<1$ which does not allow the formation of PBHs in the PS formalism { applied in the case of fluctuations that enter the horizon} (SMAP adopt a top hat filter in k-space; a gaussian window function would allow PBH formation to slightly higher masses).  This mass is $M_{\rm piv}\sim   5\times10
^{12}$h$^{-1}M_\odot f_m$, where $f_m$ is the fraction of the mass within the linear mode that collapses into a black hole.  The maximum cut off mass scale $M^*$ of the resulting PBH mass functions is then $\simeq M_{\rm piv}$ having only a slight increment with the blue spectral index $n_b$ of $0.25$ dex from $n_b=1.1$ to $4$.

(ii) Formation at a fixed conformal time, referred to as FCT.  In this case the amplitude of fluctuations is taken at the end of inflation, and in SMAP we assume that nucleation bubbles of different scales, sourced by inhomogeneities in the inflaton field, are able to produce primordial black holes as the bubbles collide with each other, in the anti peaks of the energy density during reheating.  In linear theory these anti peaks distribute in the same way as overdensities and this analogy allows  to use a PS like distribution for the masses collapsing at the anti peaks of the distribution of energy density.  In this case there is no limit to the maximum PBH mass as the formation occurs during  reheating, at or slightly before the onset of radiation domination.  The pivot scale of the primordial power spectrum has the effect of inducing a sharp feature in the mass function at $M_{\rm piv}\sim f_m\rho_{\rm crit}(z_{\rm form} (2\pi/k_{\rm piv})^3)$, the total energy density contained in a sphere of the scale corresponding to $k_{\rm piv}$ at the moment of formation of the PBHs, $z_{\rm form}$, multiplied by a factor $f_m$, the fraction of mass within the horizon that is able to collapse into the PBH.  For a typical end of inflation of $a_{\rm form}\sim2\times 10^{-26}$ this mass is $M_{\rm piv}\sim 6 \times 10^{32}$h$^{-1}M_\odot f_m$.

For $M^*<M_{\rm piv}$, the generic form of $F_{\rm process}$ in Eq. \ref{eq:dndm_gen} can be fit by a power law with index $n$ and an exponential cut off at $M^*$, i.e. by a single Schechter function.    Only FCT allows the possibility of $M^*>M_{\rm piv}$ and in this case there is a first exponential cutoff at $M_{\rm piv}$ after which an additional power law with lower amplitude and exponential cut off at $M^*$ takes over.  The connection to the physics behind PBH formation lies in the relation between  the non-linear overdensity for collapse and the fraction of mass $f_m$ of the mode that reaches linear overdensity for collapse $\delta_c$, the cutoff mass scale $M^*$, and $n_b$ and the pivot scale where the spectral index turns blue, $k_{\rm piv}$.  These relationships can be found in SMAP for both formation scenarios.  

The value of the linear overdensity $\delta_c$ is allowed to vary freely over several orders of magnitude since we do not propose a direct physical mechanism that relates this linear overdensity with the physical, non-linear one that leads to collapse.  Simulations (as well as analytic studies)  have shown that the physical collapse occurs when $\delta_c^{\rm physical}\sim 0.5$ \citep[see for instance][]{Green:1999PhRvD,Niemeyer:1998PhRvL}.    

Given the lack of a physical model we consider two possible scenarios.  One where, of all the regions that satisfy a particular choice of linear density contrast, only a fraction corresponding to the ratio of matter to total density at the average time of PBH formation, $\bar\rho_m(z_{\rm form})/\rho_{\rm crit}(z_{\rm form})$ actually collapse.  They do because only these volumes satisfy the condition for the physical density contrast $\geq \delta_c^{\rm physical}$.  This  allows the density of matter in PBHs, which evolves as a matter component,  to conform the totality of DM today.  In this case, both for HC and FCT the redshift of formation (average redshift for HC) is the one where the linear overdensity reaches $\delta_c$, and the fraction of an individual linear mode that collapses is $f_m=1$.

The other scenario is that the collapse occurs in a subvolume within every single one of all the regions that have a linear overdensity of this particular $\delta_c$ value, or larger.  The subvolume $V=f_m V_H$ containing a mass $f_m M_H$, where $V_H$ and $M_H$ are the volume and mass within the horizon, is overdense enough, with $\geq \delta_c^{\rm physical}$, and therefore collapses.  In this case, for FCT the formation time of the PBH is the same as the time the linear overdensity reaches this critical value.  For HC the actual formation of the PBH takes place when the non-linear fluctuation with $\delta^{\rm physical}\sim 0.5$ enters the horizon, i.e. at an earlier time such that the scale factor of the universe was smaller than the actual scale factor at which the linear overdense mode enters the horizon.  The PS mechanism is still applicable as long as the horizon entry of both, the non-linear and linear fluctuations, occur during the same epoch of the Universe (i.e. radiation domination), and that the ratio of scale factors between non-linear collapse ($1/(1+z_{coll})$) and linear horizon entry $1/(1+z)$ is monotonic accross the time of PBH formation; this ensures that the relative amplitude of the modes are relevant for the collapse of different masses.  The redshift of collapse can be related to the redshift of entry of the linear mode by considering the ratio of the Lagrangian region that collapses into a black hole and that of the horizon, which can be obtained considering that PBHs at formation are a fraction of the energy density of the universe such that today they constitute the dark matter.  Considering that the formation takes place during radiation domination, the relation takes the form,
\begin{equation}
    1+z_{\rm coll}=f_m^{-1/2} (1+z).
\end{equation}
The redshift of formation for the HC case is calculated by averaging the redshift of horizon crossing of different masses weighted by their relative abundances.

Even though we propose no physical model for the non-linear evolution of the overdensities that will lead to the collapse of a PBH we are able to bracket the range of possibilities that lies between these two possible views, that a fraction of the volumes that satisfy the linear overdensity for collapse turn into PBHs, or that the collapse occurs in a subvolume of each and every one of the volumes that satisfy the condition for collapse.  A physical model for PBH collapse driven by primordial inhomogeneities could lie somewhere between these two cases.

\begin{figure}
    \centering
    \includegraphics[width=0.40\textwidth]{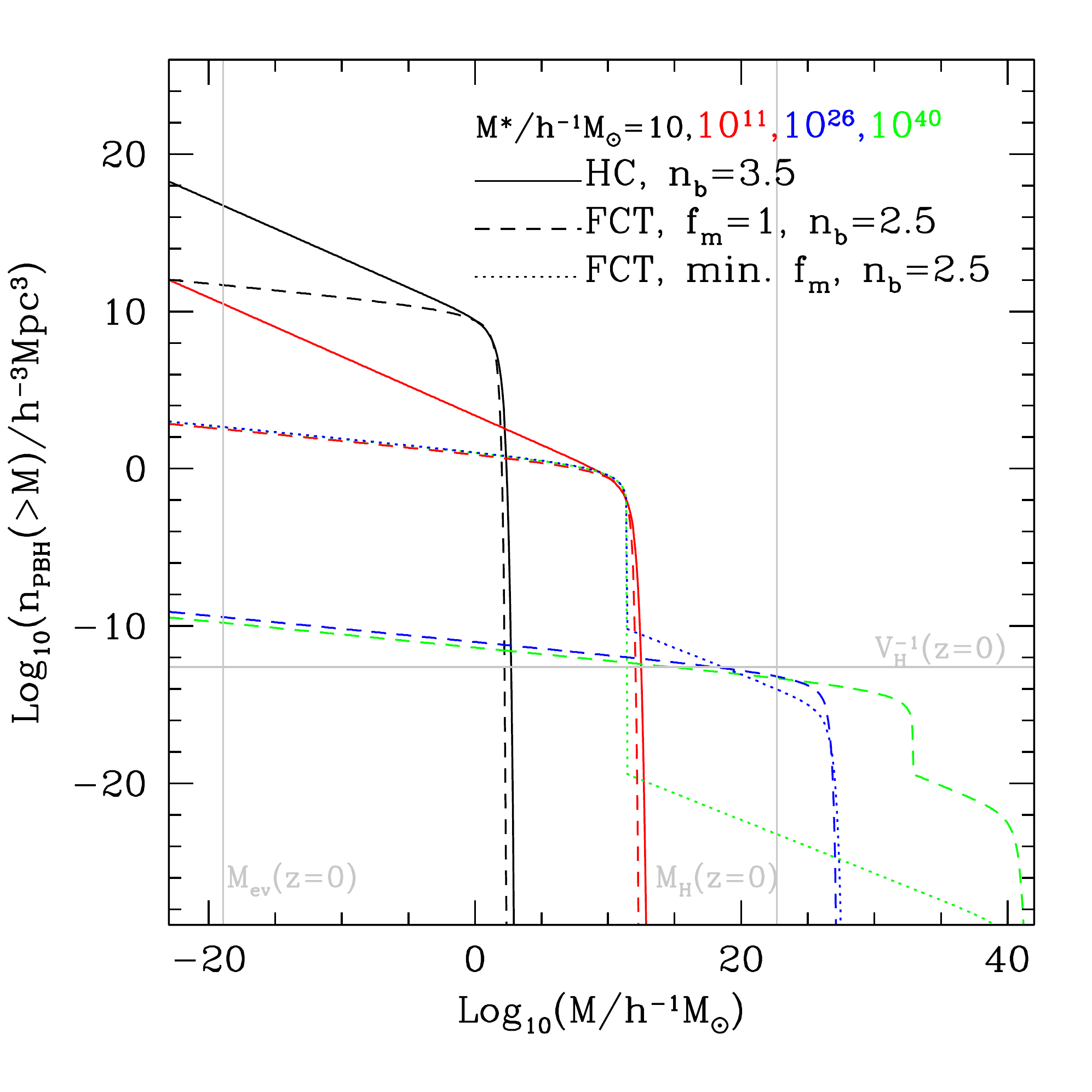}
    \caption{Example cumulative mass function of PBHs normalised such that the mass density in PBHs equals the dark matter density at $z=0$, that is, the density in objects of mass $\geq M_{ev}(z=0)$ shown by the left vertical grey line, and $\leq M_{1pH}(z=0)$, which corresponds to the intersection between the mass functions and the horizontal grey line, which represents the inverse of the Hubble volume at z=0.  HC and FCT mass functions with $f_m=1$ are shown as solid and dashed lines, respectively, and  FCT mass functions for the minimum possible $\min(f_m)$ are shown in dotted lines. Colours correspond to different characteristic masses $M^*$ indicated in the figure key.  Notice that for HC, only mass functions with $M^*<5\times 10^{12}$h$^{-1}M_\odot$, the maximum effective cut off mass for HC, are shown.  The blue spectral index is $n_b=2.5$ for FCT and $3.5$ for HC.  For FCT with $\min(f_m)$ only the higher $M^*$ cases are shown, for clarity. The grey vertical line on the right shows the mass within the $z=0$ Hubble volume. }
    \label{fig:cmf}
\end{figure}

In the remainder of this work we will mostly concentrate on the first view, that a fraction of the regions with linear overdensity $\geq \delta_c$ undergo collapse in full, i.e. with $f_m=1$, but we will also show results for $f_m<1$.  In particular, allowing for the minimum possible value  $\min(f_m)$ in HC, that is, the case of collapse of a fraction of every single horizon that satisfies the linear overdensity condition, the pivot mass decreases to $M_{\rm piv}\sim 10
^{10}$h$^{-1}M_\odot$, i.e. only by about $3$ to $4$ orders of magnitude, essentially reducing slightly the allowed range of $M^*$ for HC (i.e. the minimum value of $f_m\sim 10^{-4}$ to $10^{-3}$).  In the case of FCT, the pivot mass is much smaller than for $f_m=1$, $M_{\rm piv}\sim3\times 10
^{11}$h$^{-1}M_\odot$ corresponding to $f_m\sim 10^{-24}$. In the latter case all values of $M^*$ are permitted but the feature in the mass function due to the pivot scale of the power spectrum shifts to the smaller $M_{\rm piv}$.  At first sight a fraction of order $10^{-3}$ typical for the HC case, does not require as strong a fine tuning as the one for FCT.  However, a different model for reaheating after Inflation such that reheating ends at larger scale factors can dramatically increase $f_m$ for FCT  (e.g. \citealt{Kuro2014,Cook:2015JCAP, German:2020, Asadi:2019}) and alleviate this possible issue.  Because of this we  consider both HC and FCT scenarios in their two forms referred to generically by $f_m=1$ and $\min(f_m)$.

The slopes of the PS PBH mass distributions (at masses well below the characteristic mass) are also related to the physics of formation and to the spectral index, particularly to the blue index $n_b$.  In the FCT case the logarithmic slope  of the differential mass function $n$ is related to $n_b$ via $n=-(9-n_b)/6$. For extreme cases with $M^*>M_{\rm piv}$ in FCT, the slope changes for masses $>M_{\rm piv}$ to $n=-4/3$.  
In the HC case the relation reads $n=-(9-n_b)/4$, giving a maximum steepness for the abundance of $n=-2$ consistent with previous estimates (cf. \citealt{Carr:1975}, \citealt{Carr_2018}).

Therefore, it is possible to associate the same PBH mass function to both the HC and FCT collapse mechanisms, except that the blue spectral indices $n_b$ and linear thresholds for collapse $\delta_c$ will be different in each case.

Figure \ref{fig:cmf} shows examples of cumulative mass functions for different characteristic PBH masses for HC and FCT with $f_m=1$ (solid and dashed), and FCT with $\min(f_m)$ (dotted), with $n_b=3.5$ for HC and $2.5$ for FCT.  The left vertical grey line shows the mass of a BH that has evaporated by $z=0$ due to Hawking radiation.  No PBH should exist today below that mass if this mechanism is present in black holes.  On the other hand, the horizontal grey line shows the inverse of the Hubble volume at $z=0$.  The intersection of the mass functions with this horizontal line shows the mass of the most massive black hole with a probability of finding just $1$ such PBH within the Hubble volume (more details about this quantity can be found in the next Subsections).  One expects to find higher mass black holes in decreasing fractions of independent Hubble volumes, but we simply ignore this possibility.  As can be seen in the FCT mass functions with $f_m=1$, once the characteristic mass is higher than $M^*\sim10^{26}$h$^{-1}M_\odot$ the amplitude of the mass function is such that regardless of $M^*$ the functions look the same within the observable PBH masses at $z=0$.

Notice the feature in the $f_m=1$ FCT mass function with $M^*=10^{40}$h$^{-1}M_\odot$ at the pivot mass $M_{\rm piv}$; this is due to the change of the spectral index at this mass scale; the feature is quite sharp due to the use of a top hat window function in k-space; it would be smoother for other choices such as gaussian, or top hat in real-space.  

The feature in the $\min(f_m)$ case for FCT is at a much lower mass, $M\sim10^{11}$h$^{-1}M_\odot$, which makes most of the mass functions with characteristic masses above this value to be very similar, especially once $M^*$ is large enough that the sharp feature essentially makes the maximum mass of a PBH within the horizon to match $M_{\rm piv}$.

Also worthy of note is the lack of HC mass functions for $M^*>10^{12}$h$^{-1}M_\odot$.  This is due the condition $M^*\leq M_{\rm piv}\sim5\times10^{12}$h$^{-1}M_\odot$ in the HC formation case.

A final comment regarding the magnitude of the characteristic masses which we explored even up to $M^*=10^{40}$h$^{-1}M_\odot$ in FCT.  In no cases the masses of PBHs with abundances of at least 1 per horizon at $z=0$ is higher than $M^*=10^{23}$h$^{-1}M_\odot$ which corresponds to the mass within the $z=0$ Hubble volume (right vertical grey line).  Such PBHs are still extremely massive and, in the more extreme mass functions, all matter within the horizon lies in few PBHs.   However, we will show in later Sections (as is also shown in SMAP at least for HC) that only mass functions with $M
^*<100$h$^{-1}M_\odot$ are allowed by the different available observables that are sensitive to the abundance of PBHs.

\subsection{Effect of $f_{\rm PBH}$}

We now explore the effects of having only a fraction of the total DM mass in PBHs, $f_{\rm PBH}$.  This permits the calculation of the allowed fraction of dark matter in black holes given some particular observable that is sensitive to the presence of PBHs.

There are two ways to think about the fraction of dark matter in PBHs.  One takes into account that there is only a fraction of the mass in PBHs in, say, a halo,
$$ M_{h}^{\rm PBH} = f_{\rm PBH} M_{h} $$
and this can be used to work out the number of PBH conforming a halo, or populating a region of a given volume $V$.

The other way is to consider this fraction for the calculation of Poisson noise when thinking about overdensities.   In this case Poisson noise will contribute only in part to the realization of an overdensity, because smooth dark matter (smooth in the sense that candidate dark matter particles are usually tens of orders of magnitude lighter than PBHs) will trace $(1-f_{\rm PBH})$ of the excess mass within $V$.  The overdensities in DM coming from the initial conditions and the poisson noise will be considered independent.  In this case the contribution of Poisson noise from PBH to the total rms fluctuation reads,
$$ \sigma_{\rm Poisson}=f_{\rm PBH} \sigma_N $$
assuming that $\sigma_N$ encapsulates the Poisson noise for the patch which is being considered.

\subsection{Fraction of mass within the horizon as fluctuation becomes causal}

Here we remark on one interesting effect of the sparsity of PBHs in extended mass distributions that fall steeply with PBH mass $M$, adopting for the moment the simplified assumption that all of dark matter is in the form of PBHs, i.e. that the fraction of mass in PBHs is $f_{\rm PBH}=1$.

In these distributions PBHs of the highest masses tend to be very rare and even likely not to be found within the causal volume at early times.  Let $N(M)$ represent the cummulative number of PBHs,
\begin{equation}
    N(M)=V n(M)=V \int_M^\infty \frac{dn}{dM'}dM'.
\end{equation}
We will refer to the mass at which the cummulative number function equals one PBH per volume as $M_{1pV}$,
\begin{equation}
    n(M_{1pV})=1/V.
\end{equation}
For the special case when the volume is that of the comoving Horizon, $V=V_H(z)$, we define $M_{1pH}(z)=M_{1pV}$, which is a function of redshift because the Hubble volume increases with cosmic time, and it will enter calculations as the upper mass of PBH that are in causal contact at redshift $z$, as is the case of fluctuations that enter the horizon.

We will define the number density fraction of PBH as,
\begin{equation}
    f_{nV}(z)= \frac{\int_{M_{ev}(z)}^{M_{1pV}} \frac{dn}{dM}dM}{\bar\rho_m},
\end{equation}{}
where $\bar\rho_m$ is the average comoving density of dark matter, and the volume density fraction,
\begin{equation}
        f_{\rho V}(z)= \frac{\int_{M_{ev}(z)}^{M_{1pV}} M\frac{dn}{dM}dM}{\bar\rho_m}.
\end{equation}{}
The normalisation of the mass function $A_n$ is defined so { that $f_{\rho V_H}(z=0)=1$, where $V_H$ is the present-day Hubble volume}.

These last two quantities are also named differently for the case when $V=V_H(z)$, $f_{nH}(z)$ and $f_{\rho H}(z)$ replacing the generic volume $V$ by the Hubble volume $V_H(z)$.

\subsubsection{Evolving PBH number and mass density}

With extended PBH mass functions the content of matter in PBHs in the Universe evolves with time.  
The evolution of the average matter density, for a fraction of mass in PBH of $f_{\rm PBH}$ reads,
\begin{equation}
    \bar\rho_m(a)=( 1-f_{\rm PBH}) a^{-3} \rho_{m,0}+ \rho_{\rm PBH}(a),
    \label{eq:evmass}
\end{equation}
where, 
\begin{equation}
    \rho_{\rm PBH}(a)=f_{\rm PBH} f_{\rho H}(a) a^{-3} \rho_{m,0}.
    \label{eq:rhopbh}
\end{equation}
 
\begin{figure}
    \centering
    \includegraphics[width=0.4\textwidth]{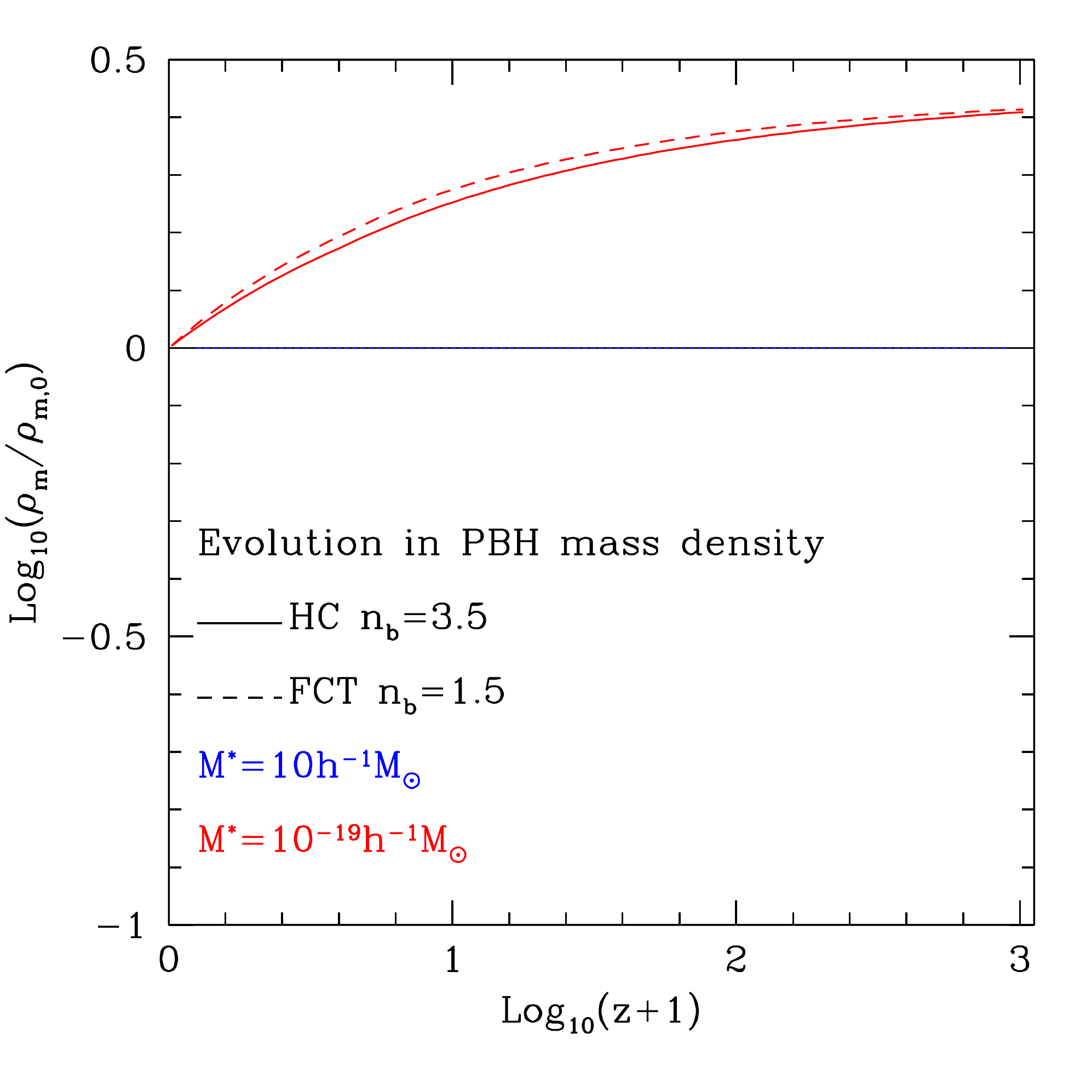}
        \vskip-.3cm
    \includegraphics[width=0.4\textwidth]{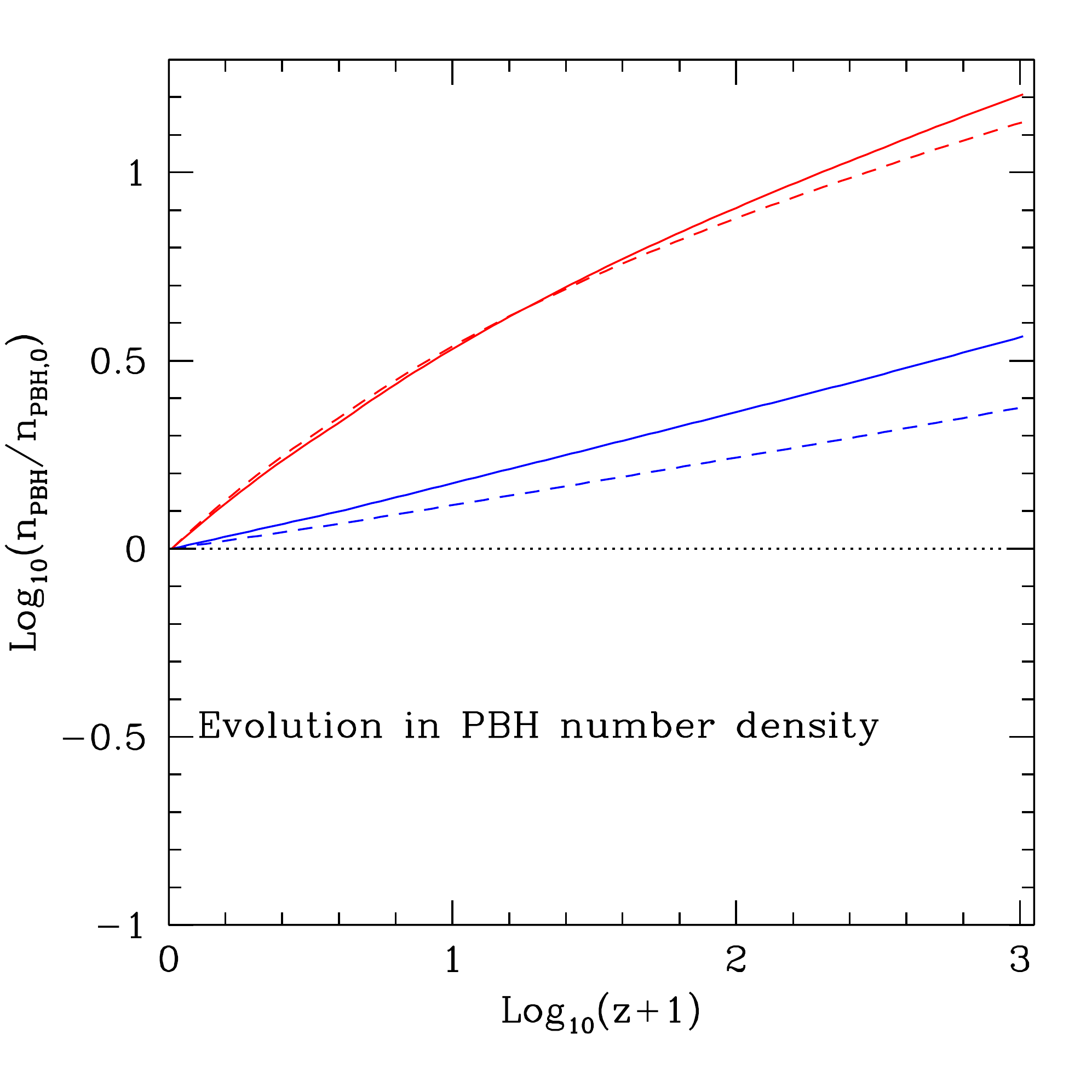}    \vskip-.3cm
    \caption{
    Top panel: evolution of the comoving mass density  as a function of redshift, for $f_m=1$ FCT and HC (differenty line types) and different values of $M^*$  (different colours, indicated in the figure key).     The horizontal dotted line represents the unit ratio (i.e., no change with time). Bottom: evolution of the comoving PBH number density as a function of redshift, normalized to their abundance at $z=0$, with same colours and lines as the top panel.  In both panels $f_{\rm PBH}=1$, i.e. all of dark matter at $z=0$ is in PBHs.}
    \label{fig:evol}
\end{figure}

We show two examples of mass and number density evolution with $f_{\rm PBH}=1$ in Fig. \ref{fig:evol} which displays the HC and FCT cases with similar overal shapes of their mass function.  The blue lines show the case of intermediate $M^*$ values which show constant density in PBHs within the horizon.  The red lines show the case of low $M^*$, close to the value of the evaporated mass of PBHs today.  The top panel shows that in this case, the evaporation causes the density of mass in PBHs to drop by more than a factor of $2$ since decoupling.  The bottom panel shows that the effect is actually stronger for the space density of PBHs.  This is due to small black holes that evaporate such that between decoupling and the present day their number density drops by more than one order of magnitude for the low $M^*$ case, and by about a factor of $3$ for the mass functions with intermediate $M^*$.  It is worthy of note that reducing the population of PBHs by two thirds does not impact the mass density in PBHs significantly, indicating that in this case most of the mass resides in higher mass PBHs. 

In the case that a fraction of the matter $f_{\rm PBH}$ is in PBHs, the evolution of the evaporated mass, and the effect of increasingly more massive PBHs entering the horizon make the total dark matter mass density to evolve with time. 
\begin{figure}
    \centering
    \includegraphics[width=0.4\textwidth]{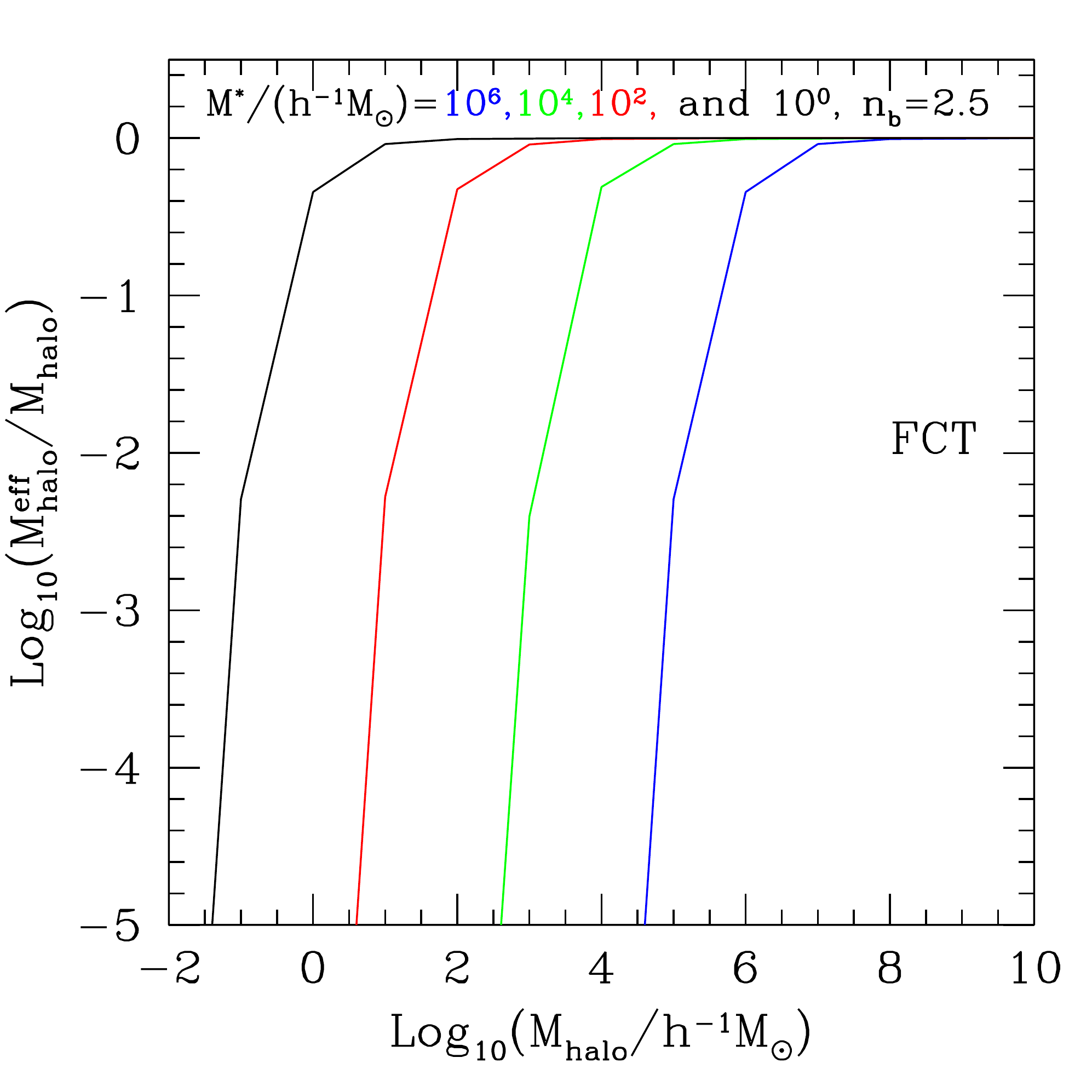}
    \vskip-.3cm
    \includegraphics[width=0.4\textwidth]{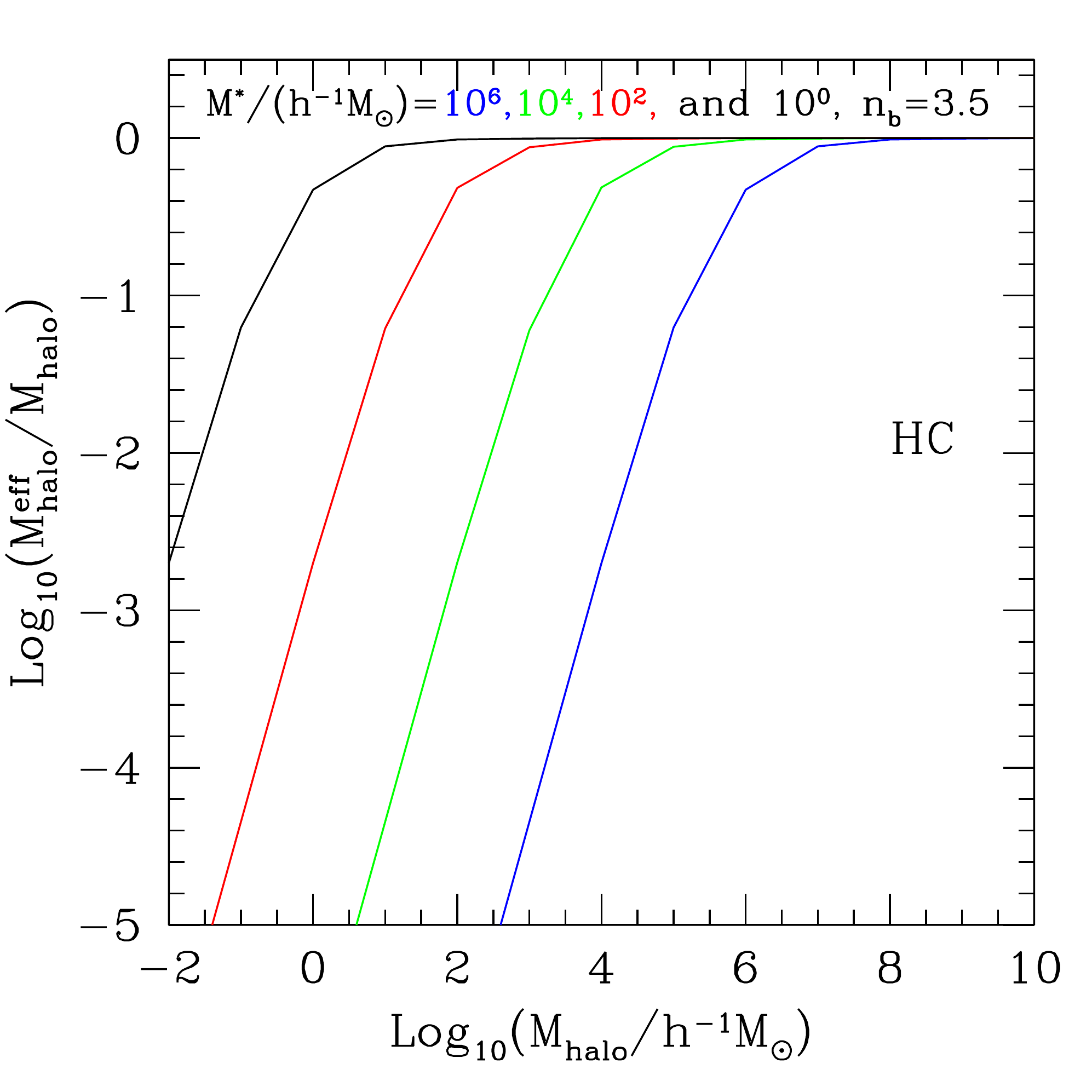}
    \vskip-.3cm
    \caption{Effective mass ratios resulting from the change in the matter density within the horizon expected for extended PBH mass functions of different characteristic mass  (different colours as indicated in the figure)  for FCT and HC (top and bottom, with $n_b=2.5$ and $3.5$, respectively). All cases correspond to growth of Poisson fluctuations since the beggining of matter domination with $f_{\rm PBH}=1$. }
    \label{fig:meff}
\end{figure}

\subsubsection{Effective halo masses}

Because in general, for extended mass distributions, $M_{1pH}<\infty$ the average mass within the horizon at redshift $z$ is,
\begin{equation}
    \bar\rho_{\rm PBH}(z)=f_{\rm PBH} f_{\rho H}(z) \bar\rho_m,
\end{equation}
which will be noticeably lower than $f_{\rm PBH}\bar\rho_m$ for all redshifts where $M_{1pH}(z)\leq M^{*}$.

Therefore, on average, because the mass contained in the overdensity is lower since $f_{\rho H}(z_{in}(M_{h}))<1$, where $z_{in}(M_{h})$ is the redshift at which $M_{h}$ enters the horizon, the effective mass of the halo that collapses from these patches will be smaller.
Taking a fraction $f_{\rm PBH}$ of the dark matter in PBH, 
$$ M_{h}^{\rm eff}=(1-f_{\rm PBH}(1-f_{\rho H}(z_{in}))) \bar\rho_m V $$
which translates into
\begin{equation}
     \frac{M_{h}^{\rm eff}}{M_{h}}=1-f_{\rm PBH}(1-f_{\rho H}(z_{in})).
\end{equation}{}
Since $M_{h}^{\rm eff}$ has the potential to be much smaller than $M_{h}$ the expected abundance of very low mass dark matter haloes can decrease by several orders of magnitude with respect to some types of CDM.  For instance, for CDM composed of particles such as the neutralino, the abundance keeps increasing slightly down to dark matter haloes of Earth mass \citep{Angulo2010}. 

The abundance of haloes can be used to place strong constraints on the parameters of PBH extended mass functions, as  Fig. \ref{fig:cmf} shows that for characteristic masses $M^*=10^{11}$h$^{-1}M_\odot$ the abundance of PBHs surviving down to $z=0$ reaches at most a value $n_{\rm PBH}\sim 100$h$^3/$Mpc$^{-3}$.  Since the Lagrangian volume of a halo of $10^{11}$h$^{-1}M_\odot$ is roughly $10^{-2}$h$^3/$Mpc$^{-3}$ this clearly shows that the dark matter in such haloes would be typically concentrated in a single PBH, and observations of dwarf satellites of nearby galaxies which are hosted by dark matter haloes of this mass,  show extended dark matter distributions (e.g. \citealt{de_Blok_2010}).

Fig. \ref{fig:meff} shows the effective masses of dark matter haloes for several values of $M^*$.  { It can be inferred} that for $M^*=10^{10}$h$^{-1}M_\odot$ the effective mass of dwarfs would be strongly affected by this effect.  Notice though that smaller effective PBH masses lower the impact on observable masses.  In particular, for the values of $M^*$ that allow a sizeable fraction of dark matter in PBHs according to SMAP, the effect becomes noticeable only at halo masses of $M_{h}\sim 10^3$h$^{-1}M_\odot$ or below, much lower than the masses of the dark matter satellites that show the possible excess with respect to observations.  

This is an approximate approach that allows to see effects coming from the sparsity of PBHs on dark matter haloes, that we will refine later on when we use the full power spectrum of density fluctuations to estimate the actual halo mass function.

\section{Poisson noise and the power spectrum}
\label{theory}

If we concentrate on measurable fluctuations, ignoring the blue part of the primordial power spectrum, the $\Lambda$CDM linear power spectrum of density fluctuations at late times, after decoupling, has the following form,
\begin{equation}
    P_{\Lambda {\rm CDM}}(k,z)=P_{\rm prim}(k) T^2(k) D_1^2(z),
    \label{eq:power}
\end{equation}
where $T(k)$ is the transfer function and $D_1(z)$, the linear growth factor.  To calculate the $\Lambda$CDM transfer function and growth factors we adopt the density parameters $\Omega_b=0.048$, $\Omega_{\rm CDM}=0.262$, $\Omega_{\Lambda}=0.69$ for baryons, cold dark matter and dark energy, respectively, and an amplitude of fluctuations on spheres of $8$h$^{-1}$Mpc of $\sigma_8=0.81$, consistent with the combined constraints from \cite{PlanckX2018}.  For the primordial power spectrum the parameters are those given in Section \ref{sec:mf}.
This power spectrum responds only to the evolution of fluctuations down to decoupling ignoring any shot noise from dark matter constituents such as primordial black holes.

If mass is distributed in discrete tracers, there is an extra component to the power spectrum coming from shot noise,
\begin{equation}
    P_{\rm Poisson}^{tracers}(k)=1/\bar n,
\end{equation}
where $\bar n$ is the average number density of tracers within the horizon, in the same units as $P(k)$.  Notice that $P(k)$ has units of $1/$Volume, i.e. a constant power indicates larger fluctuations on smaller scales.  If we consider PBHs as constituting part of the mass budget, this term will be present independently of the shape of the PBH mass distribution.

In addition to this, Poisson inhomogeneities from PBHs will also undergo the dynamics of fluctuation growth as they can affect the gravitational potential.  We will briefly review the growth of matter and Poisson fluctuations in what follows.

\begin{itemize}
    \item[(i)]When matter is in the form of fundamental particles the growth of inhomogeneities behaves as in the standard $\Lambda$CDM overdensities  during radiation domination, in the sense that they can grow since the moment they enter the horizon and all the way to z=0.  

Because of the initial conditions, the growth factor for $\Lambda$CDM matter fluctuations within the horizon is logarithmic, much slower than outside the horizon where $D_1(a) \propto a^l$ with $l=1,2$ in the matter and radiation epochs, respectively. Inside the horizon, matter density fluctuations continue to grow. These phenomena are fully taken into account by the transfer function, which  quantifies the ratio between the actual growth of a fluctuation of scale $k$ since it entered the horizon, with respect to what it would have been able to grow had it never entered it \citep{Dodelson:2003}.  

We can therefore use the full linear $\Lambda$ Cold Dark Matter transfer function $T(k)$ to go back and forth between the amplitude of $\Lambda$CDM fluctuations at the moment a given $k$-mode enters the horizon and at any later redshift.

In addition to this one also needs $D_1$ to relate the amplitude of fluctuations $P(k)$ at any redshift with the amplitude when a scale $k$ entered the horizon by finding the redshift $z_{in}$ such that the wavenumber $k$ enters the horizon, $k=2\pi/r_H(z_{in})$ where $r_H$ is the horizon size as a function of redshift. 

Since for a wavenumber that is well outside the horizon at equality, eg. $k_{outside}=0.001 $h$/$Mpc, $T(k_{outside})=1$, we can obtain the amplitude at horizon entry and use it to infer the amplitude of the fluctuations at redshift $z$,
\begin{equation}
    G_{\Lambda \rm{CDM}}(k,z)=T(k)\frac{D_1(z)}{D_1(z_{in}(k))}.
\end{equation}

\item[(ii)] Poisson fluctuations in the density of PBHs are also able to survive until matter domination because PBHs are massive enough that the absorption of radiation has no effect on them in practice, especially on their spatial distribution.  It is therefore reasonable to consider poisson fluctuations in their distribution at the end of radiation domination.

Poisson fluctuations can only affect the potential around the time when matter domination begins.  One should not consider that the Poisson fluctuation of the PBH density can grow before the time when the PBHs themselves dominate the energy budget. Poisson fluctuations in energy density are a sampling error relative to the total available energy density within a sampling volume, and if a component does not dominate the energy budget, the sampling error of counting a few more or a few less particles of that component will not have any significant effect on the energy sampled within the volume. On the other hand, the usual fluctuations of the matter component including PBHs do evolve in time, as they are not the result of a sampling of the energy within a volume, but are simply the result of the dynamics of that component, which are entirely determined by the evolution in the dominant component. Thus, even though dark matter fluctuations evolve in the usual way since their formation, the Poisson fluctuations associated with that component only evolve since matter-radiation equality.

Then, in this case the growth of fluctuations is different,
\begin{equation}
    G_{\rm Poisson}(k,z)=\max\left(1,\frac{D_1(z)}{D_1(\min(z_{eq},z_{in}(k))}\right),
    \label{Eq:growth}
\end{equation}
where the redshift to be taken into account as the beginning of the growth of the mode is whatever happens last, the redshift when the mode enters the horizon, or the redshift of matter and radiation equality \citep{Carr_2018}.
\end{itemize}
In order to obtain the correct amplitude for the Poisson power spectrum we need to take into account the growth of fluctuations in the applicable scales by using $G_i$,
\begin{equation}
    P(k)_{\rm Poisson}^{\rm PBH}(k,z)=\left(\frac{1}{\bar n(z_{in}(k))}+T_M^2(k,z)\right)G_{\rm i}^2(k,z),
    \label{eq:poisson}
\end{equation}
where the subindex $i$ corresponds to "Poisson" or "$\Lambda$CDM" depending on the choice of growth for the Poisson fluctuations.  Note that by choosing "Poisson" we may be slightly underestimating the growth of Poisson fluctuations, as  PBHs become a non-negligible fraction of the energy budget slightly before equality.
This last equation includes both the evolving and unevolving Poisson contributions, and can be used to work out present day modifications to the power spectrum from overdensities other than the $\Lambda$CDM initial conditions; it also adds another term, $T_M(k,z)$ to allow the possibility of additional phenomenology associated to the discreteness of PBHs from extended mass distributions.  It is also worth to note that the evolution of the number density of black holes due to evaporation and to their sparsity, characteristic of extended PBH mass functions and entirely absent in monochromatic ones, can introduce a dependence of the Poisson noise on scale, since different scales enter the horizon at different cosmic times, and therefore they can do so with different number densities and Poisson fluctuations.

{The effect of the fraction of mass in PBHs is such that it multiplies directly the variance. The total $P(k)$ is then,
\begin{equation}
    P(k,z)=P_{\Lambda {\rm CDM}}(k,z)+f_{\rm PBH}^2 P^{\rm PBH}_{\rm Poisson}(k,z).
\end{equation}{}

\subsection{Monochromatic PBH mass distributions}

In the case of monochromatic distributions, PBHs have a unique mass and the Poisson component is $P^{\rm PBH}_{\rm Poisson}(k,z)=(1/\bar n(z_{in}(k)))  G_{\rm Poisson}(k,z)^2$, where the comoving number { density of PBHs is constant for monochromatic PBHs, $\bar n(z_{in}(k))=\bar n$}.

Including the effect of the fraction of mass in PBHs we obtain,
\begin{equation}
    P_{\rm Poisson}^{\rm PBH}(k,z)=\frac{f_{\rm PBH}^2}{\bar n}  G_{\rm Poisson}(k,z)^2.
\end{equation}
This equation ignores the limit where a mode $k$ contains $\leq 1$ black hole.  There would be a single PBH of $\sim10^{13}$h$^{-1}M_\odot$ within the equivalent real space volume of  $k_{\rm NL}=1$hMpc$^{-1}$.  Therefore if the constrained fraction of DM in PBHs of this mass (or higher) is close to $f_{\rm PBH}=1$ at these wavenumbers, our calculations need to be revised to include the change in the $\Lambda$CDM power spectrum due to the lack of DM on these and larger wavenumbers.
}

\begin{figure}
    \centering
    \includegraphics[width=0.4\textwidth]{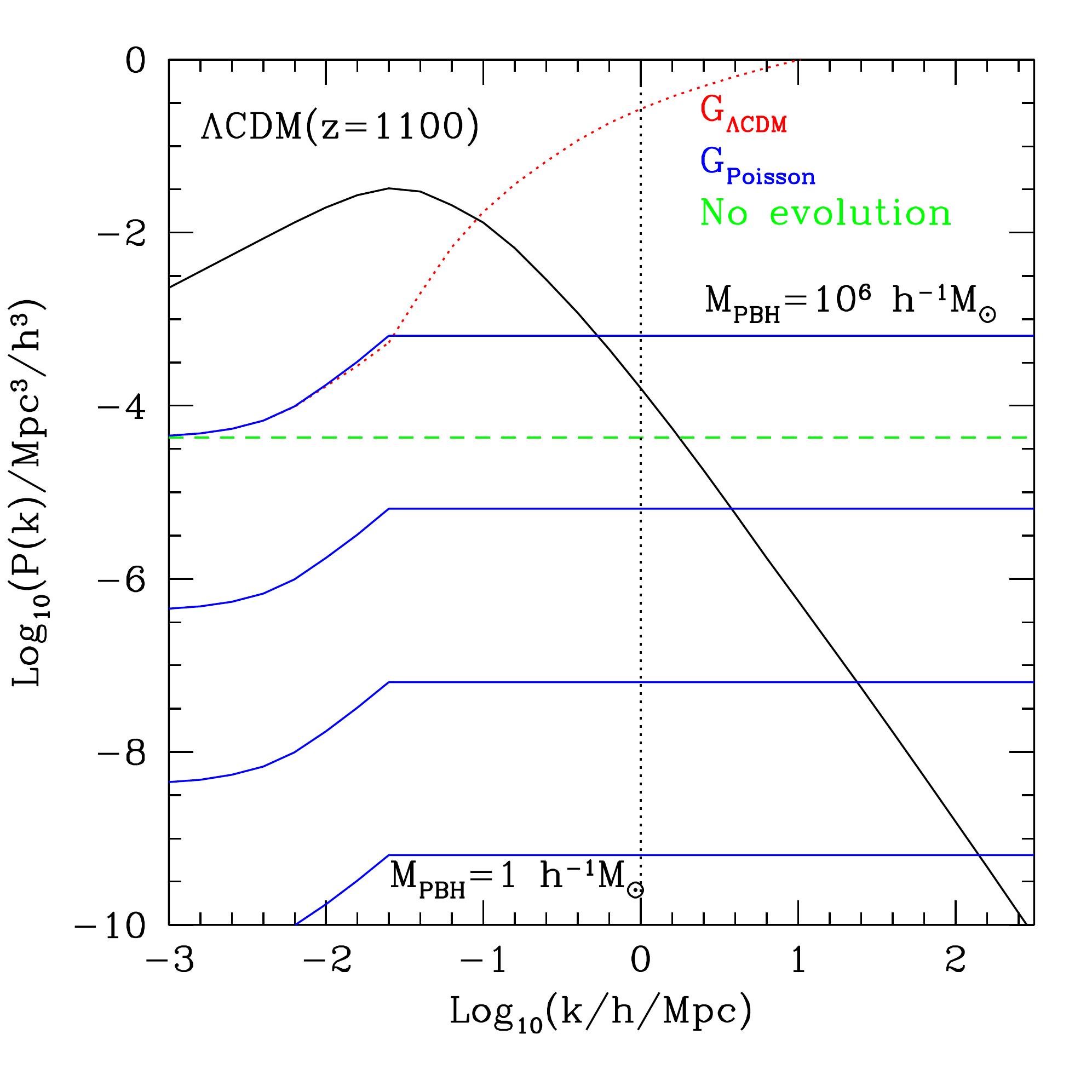}
    \vskip-.3cm
    \caption{Power spectra from Poisson noise for monochromatic PBH mass functions of masses $10^{0}$ to $10^6$ in increments of $2$dex (bottom to top blue solid lines) for growth since equality.  The dotted line shows the case of $M_{\rm PBH}=10^6$h$^{-1}M_\odot$ for growth since horizon entry, whereas the dashed is the case of no growth.  For comparison, the black solid line shows the $\Lambda$CDM power spectrum at $z_{CMB}=1100$. The vertical dotted line shows the standard wavenumber up to which which we require the Poisson power spectrum not to exceed $10\%$ of the value of the $\Lambda$CDM power spectrum.}
    \label{fig:pkpmon}
\end{figure}

Figure \ref{fig:pkpmon} shows examples of Poisson $P(k)$ for different monochromatic PBH masses and different cases of evolution of the Poisson fluctuations.  The case with no evolution of the dashed line appears as a constant line as the power spectrum is shown per unit volume which makes the Poisson spectrum simply the inverse of the number density of matter tracers, i.e. of PBHs.  The blue solid lines show an increment from large to small scale (small to large wavenumber) up to the scale of the horizon at the time matter domination starts, since this case shows the evolution since equality.  For comparison, the red dotted lines show growth since the mode enters the horizon, which exceeds the expected growth for Poisson fluctuations.  As can be seen the Poisson power can overcome the $\Lambda$CDM spectrum (shown in black) depending on the assumed evolution of Poisson fluctuations and the PBH mass; we use this to estimate a maximum allowed fraction of mass in PBHs with Eq. \ref{eq:fpkmon}.

\begin{figure}
    \centering
    \includegraphics[width=0.4\textwidth]{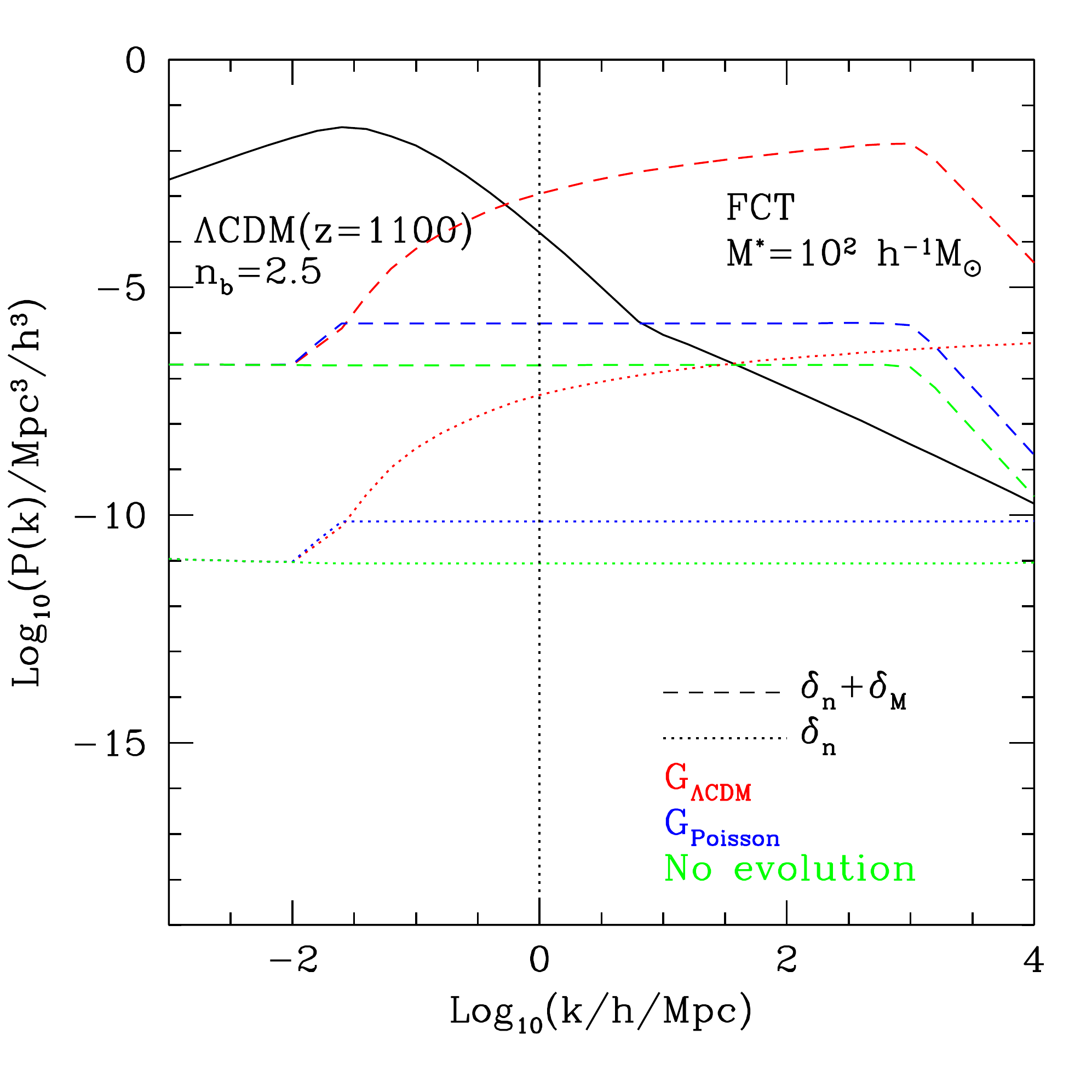}
    \vskip-.5cm
    \includegraphics[width=0.4\textwidth]{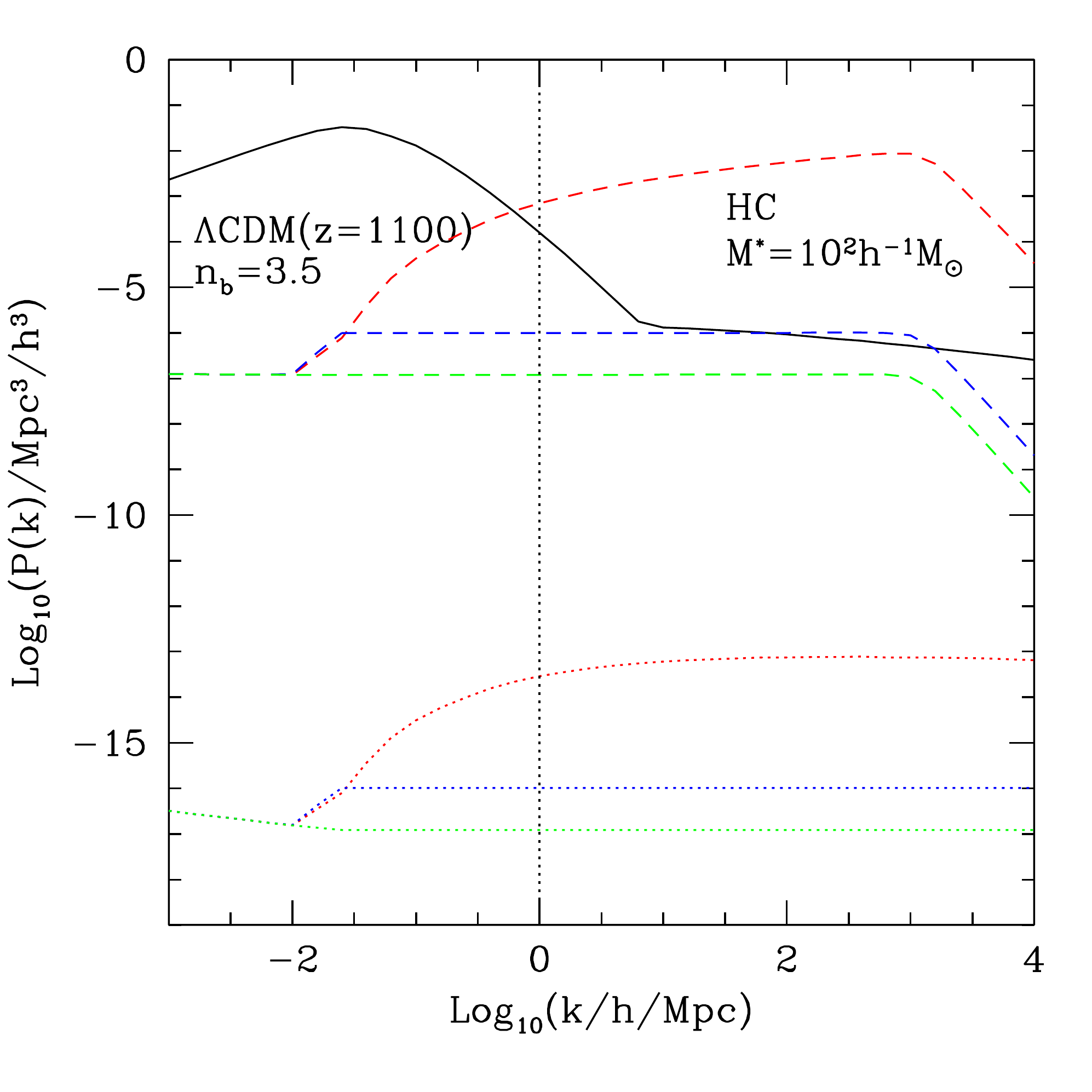}
    \vskip-.3cm
    \caption{Power spectra at $z=1100$ from Poisson noise for FCT and HC PBH mass functions (top and bottom, respectively) of characteristic mass $M^*=10^{2}$h$^{-1}M_\odot$ and $n_b=2.5$ and $3.5$ (for FCT and HC), including different terms, the one due to the number density alone in dotted, and including also noise in the total matter due to Poisson acting on the extended mass functions in dashed, with no evolution in green, evolution since the modes enter the horizon in red, and evolution since matter domination in blue.    For comparison, the black solid line shows the $\Lambda$CDM power spectrum at the same redshift; the break at $k_{\rm piv}=10$Mpc$^{-1}$ shows the effect of the blue index in the primordial power spectrum.  The vertical dotted line marks the standard wavenumber used here to calculate allowed fractions of mass in PBHs. }
    \label{fig:pk}
\end{figure}

\subsection{Extended mass distributions}
\label{ssec:pkdndm}
The variance in the matter density of PBHs within a volume corresponding to the wavenumber $k$ is related to the actual shape of the extended mass distribution $dn/dM$.

We begin by illustrating how the relative fluctuation in number density and in the PBH masses enter the relative fluctuation of the mass density.  In a simplified but illustrative picture one can write,
\begin{equation}
\sigma^2(\rho_m)=\sigma^2(n)\left<M\right>^2+\sigma^2(\left<M\right>)n^2,
\label{eq:cadena}
\end{equation} 
where $n$ is space density and $\left<M\right>$ is the average mass of PBHs, and the variances are assumed independent of one another.
In particular notice that if the distribution is monochromatic, the fluctuation of the average mass of PBHs vanishes and we are left with the Poisson term  $\sigma(\rho_m)=n^{1/2}\left<M\right>$.  In what follows we  present both terms of the fluctuation for the case of extended PBH mass distributions, so as to calculate the full fluctuation in the PBH mass density due to Poisson noise.  

Let us define the density of matter in PBHs, $$\bar\rho_{\rm PBH}(V,z)=f_{\rm PBH}f_{\rho V}(z)\bar\rho_m=\int_{M_{ev}(z)}^{M_{1pV}}M dn/dM dM,$$ and define the number density of PBH within volume V,
$$\bar n(V,z)=\int_{M_{ev}(z)}^{M_{1pV}}dn/dM dM.$$

The fluctuation in the mass due to Poisson noise can be approximated by 
\begin{equation}
    \delta_M^2(V,z)=V \left( \frac{1}{\bar\rho_{\rm PBH}}\int_{M_{ev}(z)}^{M_{1pV}} M' \frac{dn}{dM}(M') \Delta^2_{N_{\delta M}} dM'\right)^{-1},
    \label{eq:sigmamm}
\end{equation}
where $\Delta^2_{N_{\delta M}}=\max(1,N_{\delta M})$, to avoid relative fluctuations above unit in a given mass bin and, $$N_{\delta M}=V \int_{M'}^{M'+\delta M}dM'' \frac{dn}{dM}(M'').$$ When numerically integrating this equation, we set $\delta_M$ to the integration step in mass, which is chosen for convergence.  

Combining the Poisson noise in number and in mass, and including the effect of fraction of mass in PBHs which directly multiplies  the variance, 

\begin{eqnarray}
    \label{Eq:pk}
    P_{\rm Poisson}^{\rm PBH}(k,z)& = & f_{\rm PBH}^2 G_{\rm Poisson}(k,z)^2  \nonumber \\  & \times& \left( \frac1{\bar n(V_k,z_k)} +\delta_M^2(V_k,z_k)\right), 
\end{eqnarray}
where $z_k$ is the redshift when the mode $k$ starts to grow, set to equality -- or later for larger modes -- in our main analysis, and $V_k=(2\pi/k)^3$ is the volume associated to the mode $k$.  The terms between parenthesis correspond to the terms  of Eq. \ref{eq:cadena} which was presented to illustrate the Poisson noise in number density and mass; in particular, the second term corresponds to $T_M(k,z)$ from Eq. \ref{eq:poisson}.
Note that the evolving Poisson power spectrum depends on the redshift the mode $k$ either enters the horizon or starts to dominate the potential, $z_k$, as well as the redshift at which the power spectrum is calculated, $z$.

Fig \ref{fig:pk} shows the resulting power spectra for the poisson component of dark matter composed entirely of PBHs for the FCT and HC scenarios of extended PBH mass distributions, for the different cases of growth of poisson fluctuations after equality of matter and radiation, and since the modes enter the horizon, compared to the $\Lambda$CDM 
power spectrum, all at $z=1100$.  As can be seen, the poisson spectrum is able to dominate over the $\Lambda$CDM case at small scales, depending on the assumed growth.  

The common feature is that there is a dependence on scale of the Poisson variations in number and in mass, which result from a dependence of the scatter in number and mass within the horizon with time, i.e., as a function of the volume associated to the mode and the time when the Poisson fluctuation becomes dominant.  The dependence arises from there being only PBHs with at least the mass of evaporated black holes $M_{ev}(z)$ at a given redshift $z$, and with at most the mass that results in an abundance of just $1$ PBH per volume, $M_{1pV}$; higher mass PBHs will be present in decreasing fractions of spheres of sizes larger than those corresponding to the mode $k$.  This latter detail is in part where the Poisson effect on mass density of PBHs comes from, aside from there being noise in the mass of PBHs that are within the horizon.

What we see in the figure is that, in the case of no evolution (green lines), the Poisson power in number density decreases with wavenumber for small $k$, which indicates that the variations of $M_{1pV}$ and $M_{ev}$ combined result in higher number densities of PBH as the redshift increases (very slightly for FCT but clearly noticeable for HC).  When the variation in mass is added in the no evolution case the noise is orders of magnitude higher and roughly constant except at very high wavenumbers where the noise decreases, showing that $M_{1pV}$ is low enough that the abundance of PBHs rapidly increases to high numbers lowering the importance of the mass component of the noise.  

It is also interesting to note that the slope of the PBH mass function has a strong influence on the amplitude of the Poisson amplitudes in number and mass.  The variance in number is lower in HC than in FCT, but the variance in mass is higher, with an effect of $10$ orders of magnitude increase for HC compared to only $\sim 2$ for the FCT mass function; this is simply due to the steepness of the mass function; a steeper drop of the mass function increases the importance of low mass PBHs, thereby lowering the mass component of the Poisson fluctuation.  

Any appreciable changes to the power spectrum of density fluctuations at $k=1$hMpc$^{-1}$ can  likely be observed today as this affects  structures on scales of clusters of galaxies; it is difficult to disentangle non-linear effects at these scales but it is safe to say that at present we do not expect orders of magnitude differences with respect to the $\Lambda$CDM predictions.  Therefore, the condition in the extended mass function case is the same as the one for the monochromatic distribution of Eq. \ref{eq:fpkmon} that at this scale the Poisson contribution is at most $10\%$ that of $\Lambda$CDM at $k_{\rm NL}$ or smaller wavenumbers.

\section{Poisson noise and the abundance and clustering of dark matter haloes}
\label{dndm}

Dark matter haloes result from the evolution of { overdensities in the matter density field which exceed the linear threshold for collapse.
In some cases the population of PBHs will be able to track these overdensities in an accurate way and the halo mass function will correspond to $\Lambda$CDM; the reason for this is that we consider PBH seeds to be randomly distributed in space  in the Lagrangian volumes of haloes prior to their horizon entry after which primordial overdensities are modified into $\Lambda$CDM shape.  However, shot noise fluctuations can also contribute to the linear overdensity and produce DM haloes following a Poisson rather than $\Lambda$CDM spectrum.  Additionally, as the clustering of haloes depends on their rarity (rarer density peaks cluster more strongly in gaussian random fields, see for instance \citealt{Mo:1996}), a modification to the halo mass function can also be accompanied by a change in the clustering of haloes as a function of halo mass.  

It should be kept in mind that in both cases, the relevant fluctuations are the linear ones, which opens the possibility to explore the small effects introduced by Poisson noise from PBHs on scales that are difficult to study using the measured power spectrum at the redshifts where halo mass function and clustering are measured ($z\sim0$)\footnote{Simulations show that this is not strictly true as the halo mass function and clustering measured in simulations is only fit when non-linear corrections are introduced (e.g. \citealt{SMT2001}).  The departures would again need to be stronger than those introduced by non-linearities to be able to detect/reject PBHs as the source of change in abundances and clustering of haloes. }.  Both baryon effects (e.g. \citealt{Sawala_2013}) and primordial non-gaussianity (e.g. \citealt{Mana2013}) also change the abundance of massive dark-matter haloes but are not taken into account in this work.

We} will explore these two effects separately, the changes to the mass function and clustering of haloes due to PBH Poisson fluctuations, in the next subsections.

\begin{figure}
    \centering
    \includegraphics[width=0.37\textwidth]{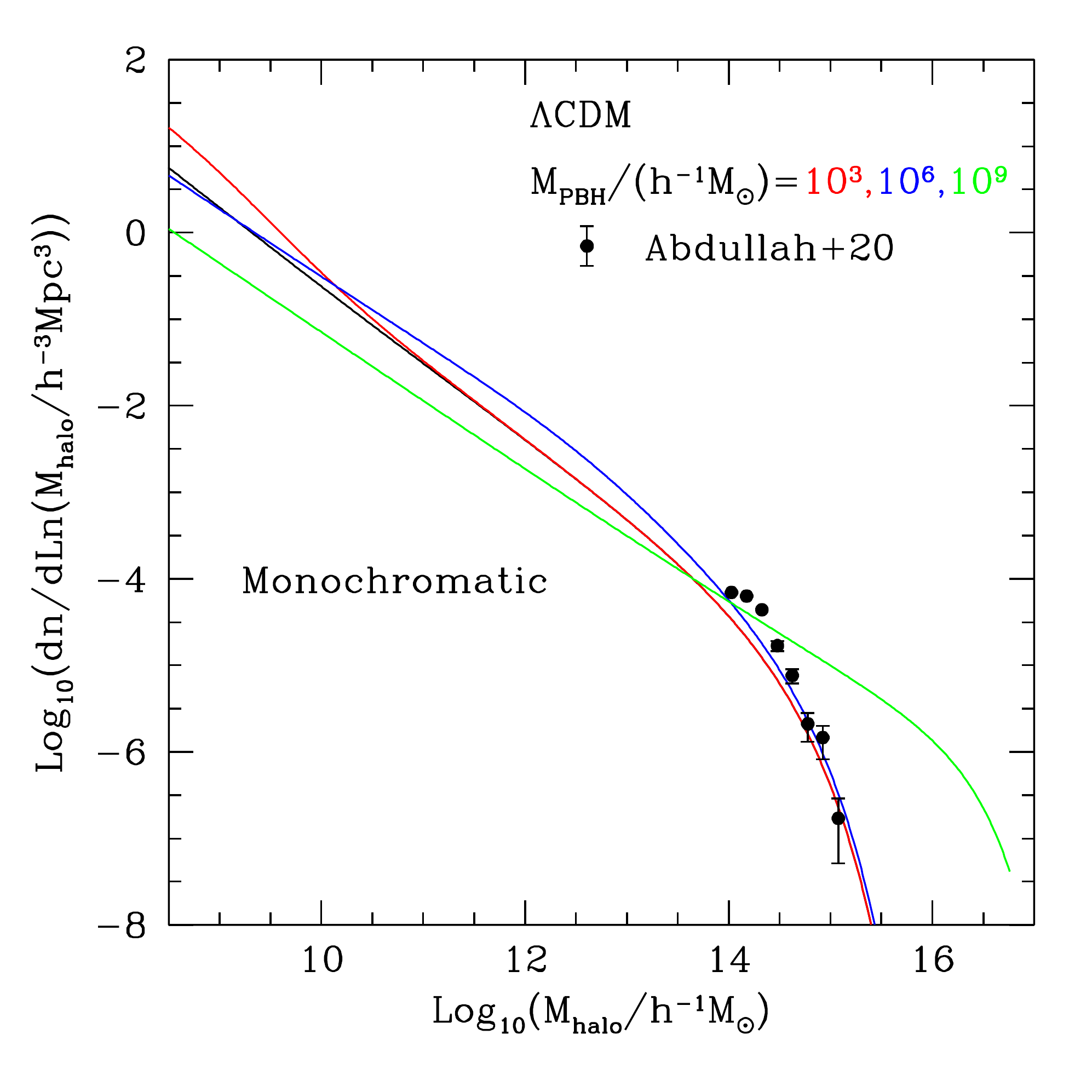}
    \vskip-.6cm
    \includegraphics[width=0.37\textwidth]{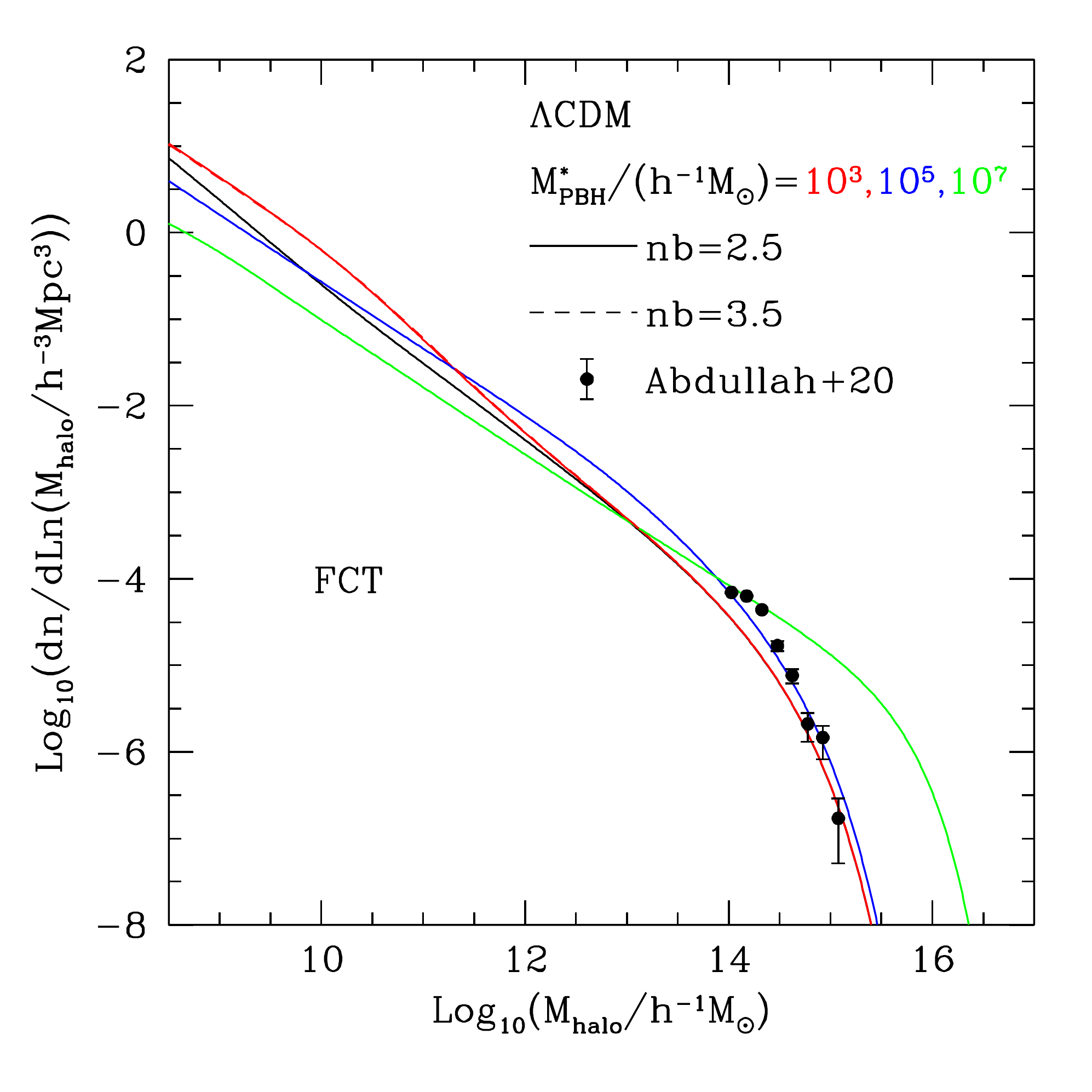}
    \vskip-.6cm
    \includegraphics[width=0.37\textwidth]{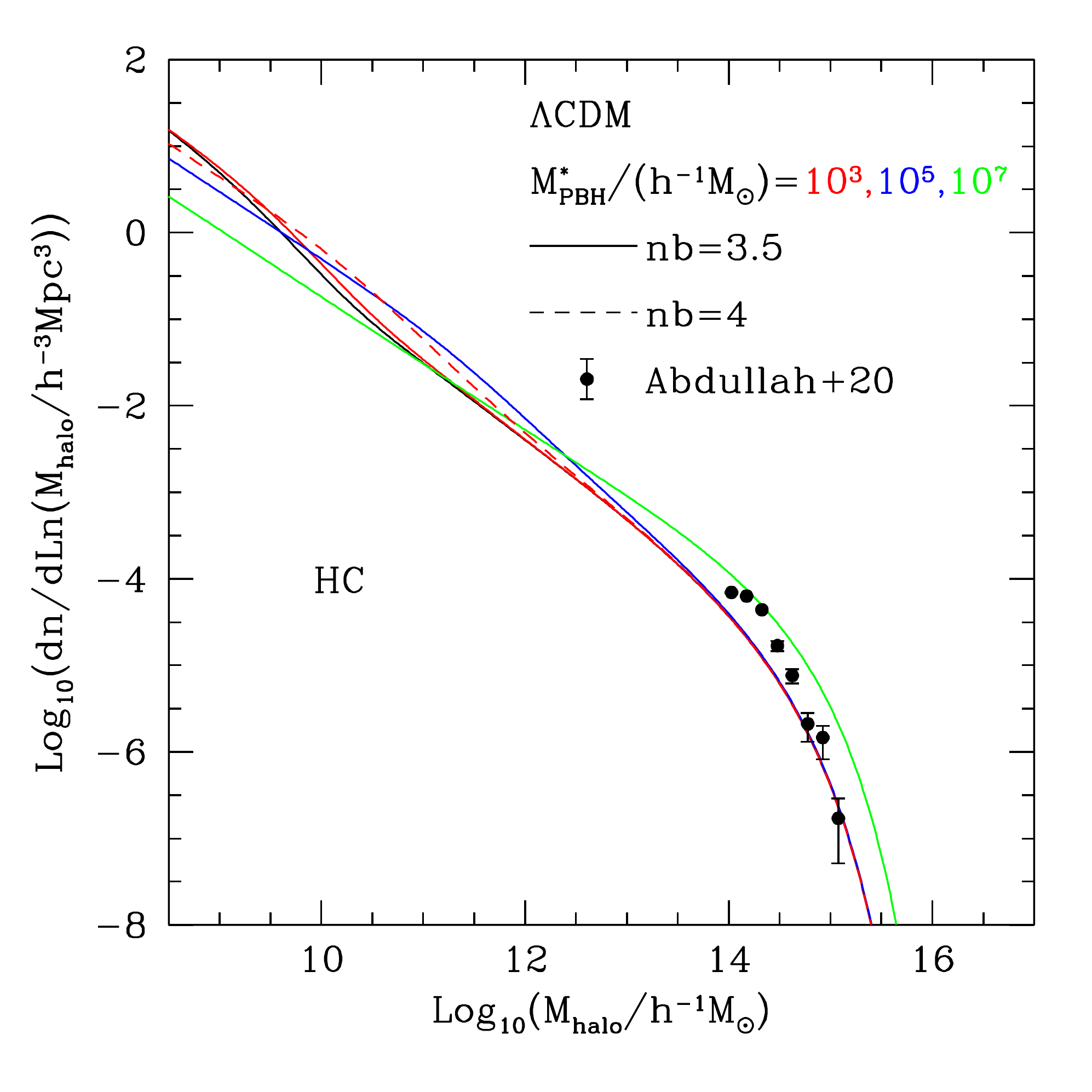}
    \vskip-.5cm
    \caption{Halo mass functions resulting from the full power spectrum consisting of $\Lambda$CDM fluctuations and Poisson noise from monochromatic (top) and FCT and HC extended PBH mass functions (middle and bottom panels, respectively).  Colours correspond to different PBH masses or characteristic masses  (as indicated in the figure) and in the case of extended distributions, different line types show results for different blue spectral indices. All cases correspond to growth of Poisson fluctuations since the beginning of matter domination. The black solid lines in all panels show the pure $\Lambda$CDM halo mass function, as reference, and the filled circles with errorbars show a recent measurement of the halo mass function by \protect\cite{Abdullah2020}.}
    \label{fig:hmf}
\end{figure}

\begin{figure}
    \centering
    \includegraphics[width=0.38\textwidth]{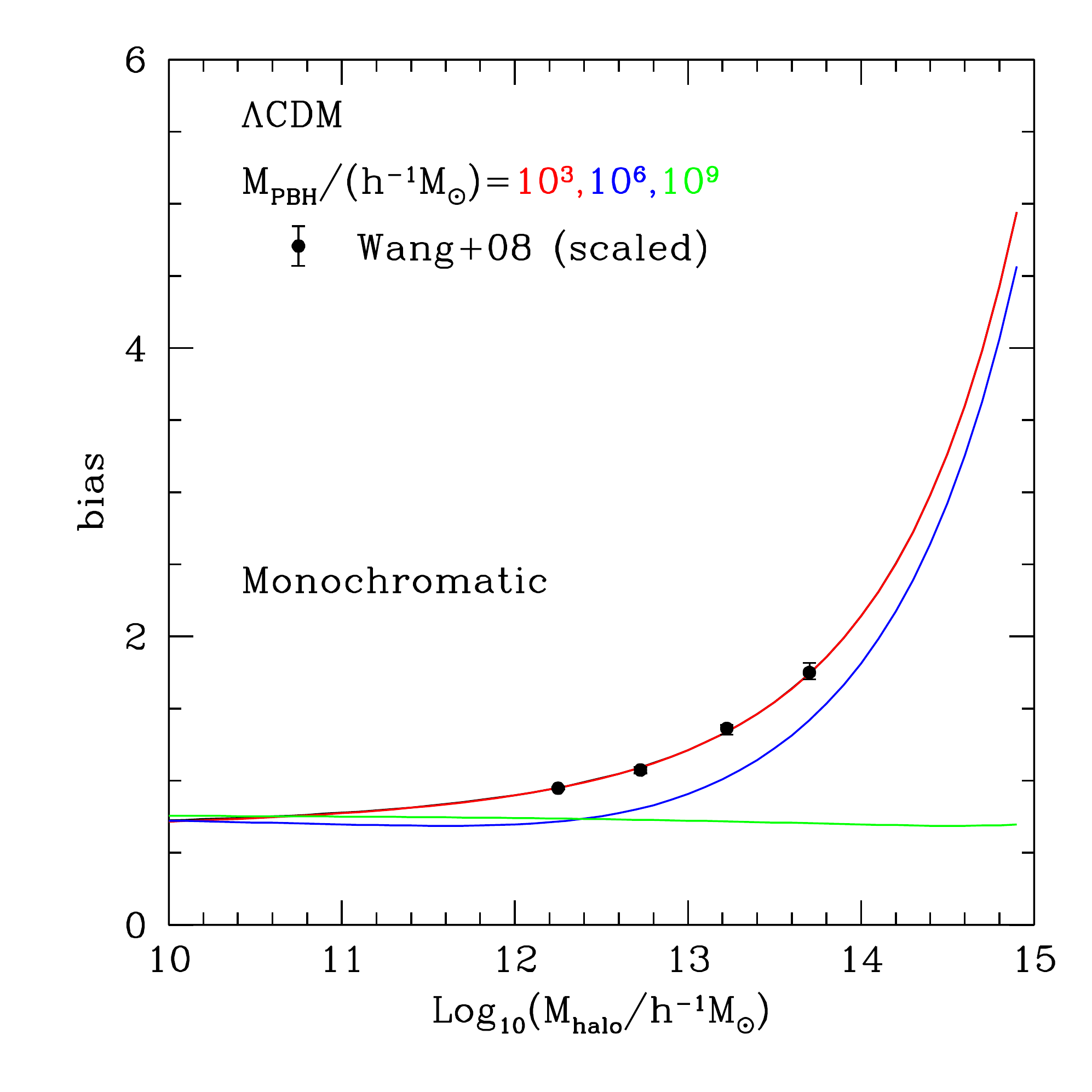}
    \vskip-.5cm
    \includegraphics[width=0.38\textwidth]{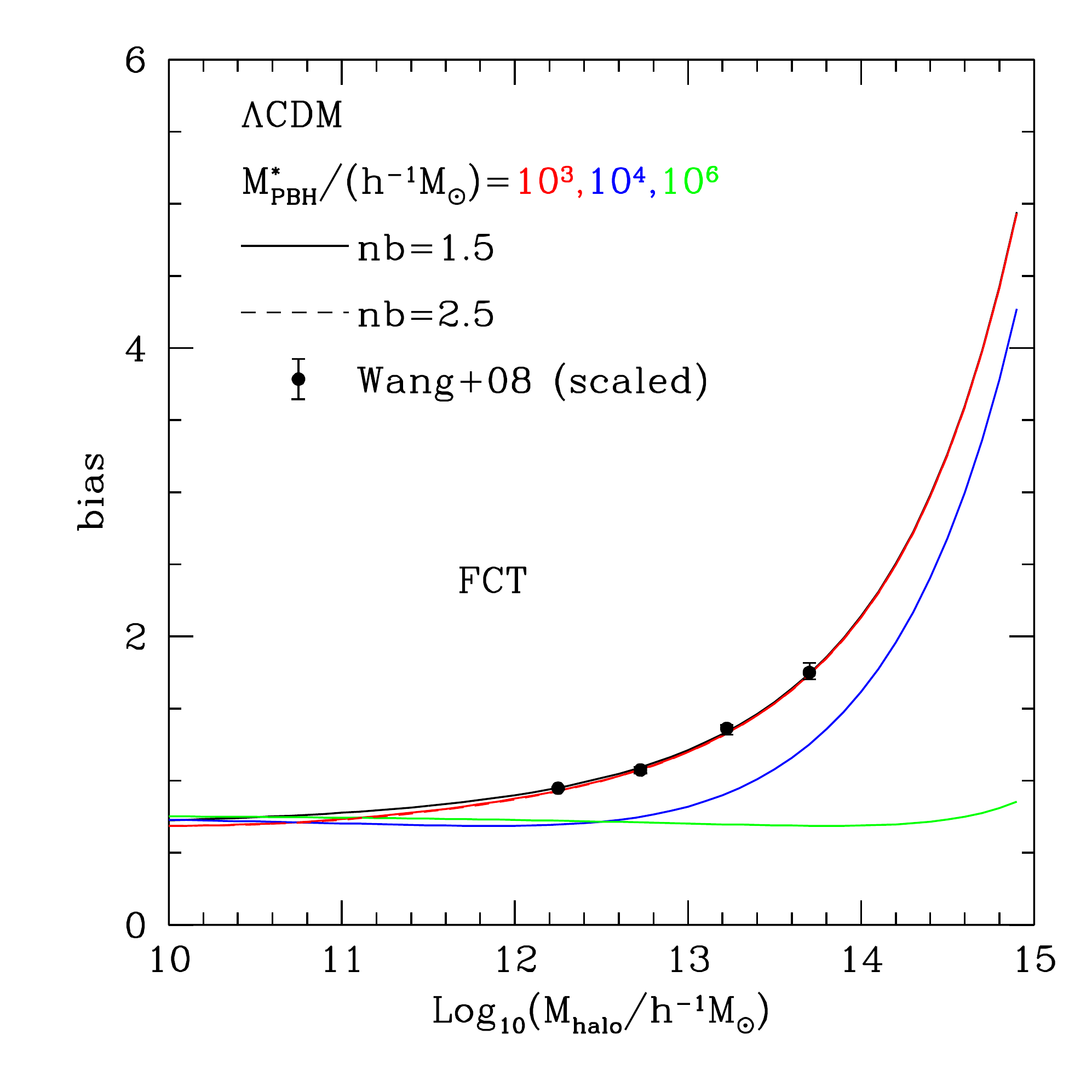}
    \vskip-.5cm
    \includegraphics[width=0.38\textwidth]{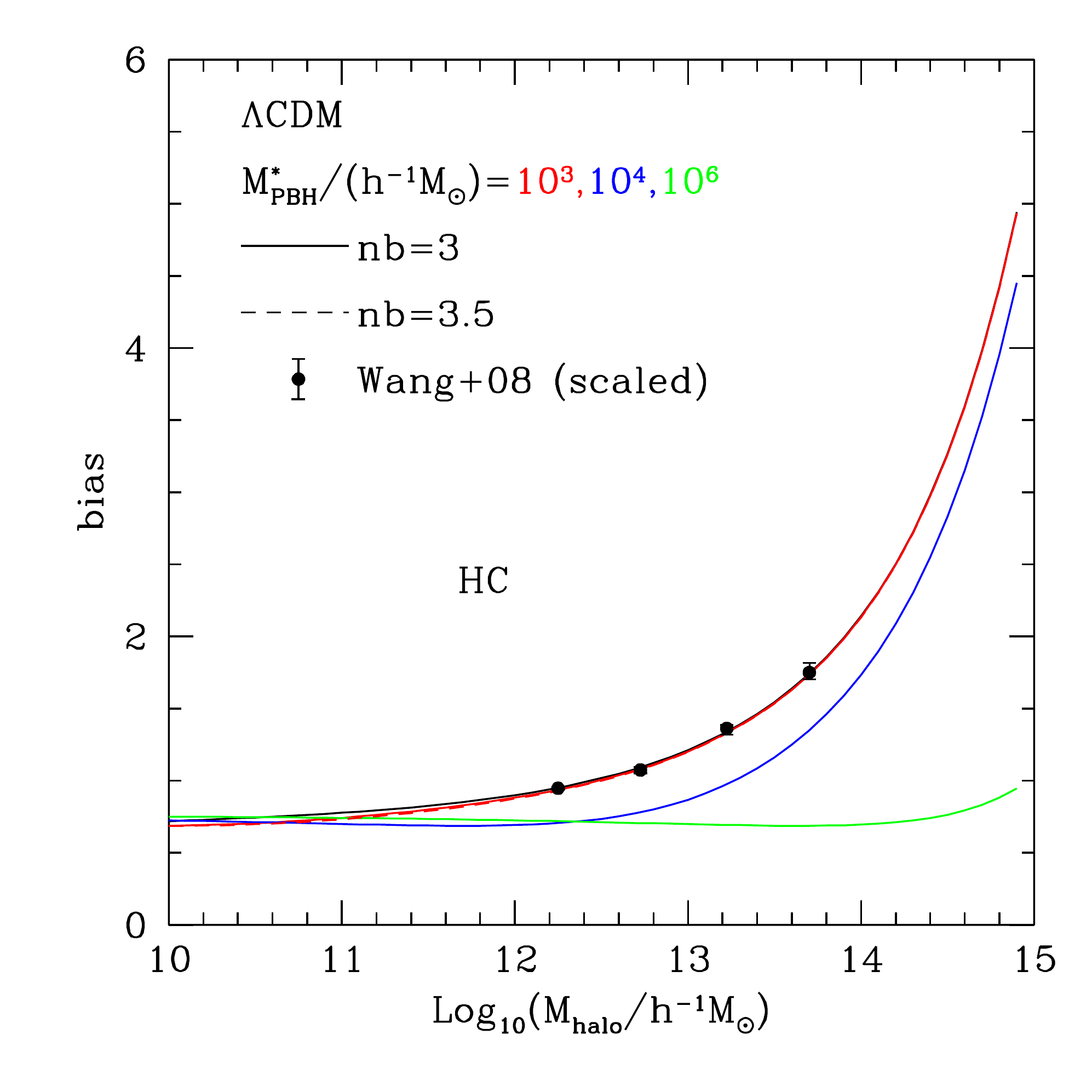}
    \vskip-.3cm
    \caption{Halo bias resulting from the full power spectrum consisting of $\Lambda$CDM fluctuations and Poisson noise from monochromatic (top) and FCT and HC extended PBH mass functions (middle and bottom, respectively).  Different colours show different PBH masses (monochromatic) and characteristic masses  (extended distributions).  For the latter, line types correspond to different blue spectral indices. All cases correspond to growth of Poisson fluctuations since the beginning of matter domination. The black solid lines in all panels show the pure $\Lambda$CDM halo mass function, as reference.  The solid points with errorbars show the measurements by \protect\cite{Wang2008}, scaled to the amplitude of the $\Lambda$CDM bias for the lowest mass bin.}
    \label{fig:bias}
\end{figure}

\subsection{Halo mass function}

The actual mass function of dark matter haloes can be calculated using the same general approach presented in Section \ref{sec:mf}, except that the collapse of haloes takes place during matter domination, in  all patches with linear overdensities higher than a given threshold for collapse, $\delta_c^h$ \citep{Mo:1996}.  We apply this formalism to find out the actual present day halo mass functions resulting from adding the Poisson noise to the $\Lambda$CDM power spectrum of density fluctuations.  To do this we use the publicly availabe code by E. Komatsu\footnote{\url{https://wwwmpa.mpa-garching.mpg.de/~komatsu/codes.html}} to calculate the mass function using the power spectra with the contribution of Poisson fluctuations at $z=1100$ (some of which are shown in Figs. \ref{fig:pkpmon} and \ref{fig:pk}) evolved to $z=0$.  

Fig. \ref{fig:hmf} shows that the Poisson contribution can produce important and noticeable changes in the halo mass function with respect to the $\Lambda$CDM model (lines, colours indicated in the figure key), irrespective of the PBH mass function.  The points show recent measurements from \cite{Abdullah2020}, who use an updated group catalogue from the Sloan Digital Sky Survey-Data Release 13  \citep{Albareti2017}, to estimate masses using the abundance matching technique as in \cite{Wang2008}.  

For the monochromatic case, the PBH masses we chose to show are within the range that are currently ruled out by different observations such as lensing \citep{Tisserand:2007}, x-ray binaries \citep{Inoue_2017} and large scale structure effects \citep{Carr_2018}.  However, the halo mass function does not depart strongly from the $\Lambda$CDM case even for $M_{\rm PBH}=10^6$h$^{-1}M_\odot$.  

For extended distributions, it is interesting to note that the resulting mass functions with Poisson noise from PBHs shown in the figure are in regions of the PBH mass function parameter space that precludes $100\%$ of the dark matter to be in the form of PBHs according to SMAP and, still, there are cases where the mass functions are quite similar to the $\Lambda$CDM case as well as to the observational points, such as the red dashed lines.

It is also worth mentioning that in some cases, especially those shown for HC, the abundance of low mass haloes $M_{\rm halo}=10^9$h$^{-1}M_\odot$ is suppressed relative to $\Lambda$CDM (black lines), which could affect the number of satellite subhaloes in galaxies (see \citealt{Fielder2019} for a recent update on the missing satellite problem and its possible solutions).  These particular models are apparently in good agreement with observations and $\Lambda$CDM at the high mass end.  

In the next section we will quantify which PBH mass functions provide halo mass functions that depart the most from the observations, and use this to reject regions of the parameter space of monochromatic and extended PBH mass functions.

To understand the excess of high mass haloes for high $M_{\rm PBH}$ or $M^*$ (monochromatic and extended mass functions, respectively) one can go back to the examples of power spectra of Figs. \ref{fig:pkpmon} and \ref{fig:pk}, where it is clear that the Poisson contribution increases the amplitude of fluctuations on larger scales for higher values of $M_{\rm PBH}$ and $M^*$ for each PBH mass function case.  This translates into a larger fraction of fluctuations above the threshold for halo collapse, which produces higher abundances of high mass haloes.  Incidentally, this also means that these high mass objects are in peaks of the density field that are not as rare as in the standard $\Lambda$CDM case, and that are sourced by inhomogeneities that are spatially uncorrelated.   In the  next subsection we will quantify the effect that this has on their clustering.

\subsection{Clustering of haloes constituted by primordial black holes}
\label{sec:bias}

We can also use the actual linear power spectrum that contains both the fluctuations in matter coming from primordial overdensities as well as the ones of PBH Poisson origin, evolved to the present day, to calculate the bias factor of haloes as a function of their mass.  \cite{Mo:1996} presented an analytical model for the bias factor,
\begin{equation}
    b(M,z)=1+\frac{{\nu (M)}^2-1}{\delta_c^h(z)},
\end{equation}
that shows the dependence of the bias on $\nu$ and $\delta_c^h$, the significance of peaks in the matter distribution from which the haloes form, and the linear overdensity for halo collapse.  

We use the parametric modification of the \cite{Mo:1996} presented by \cite{Sheth2001} which provides a better fit to results from numerical simulations.  Notice that this allows to obtain the bias of haloes starting from power spectra of any form.  We are aware that the bias formulae were not tested with power spectra that differ from $\Lambda$CDM as much as the ones we present here, for example for PBH mass functions with $M^*> 10^4$h$^{-1}M_\odot$, but this still allows to determine the PBH mass function parameters that produce bias functions that show some significant departure. 

To calculate the bias of haloes we use the implementation of the \cite{Sheth2001} formulae by E. Komatsu$^1$ which uses as input the linear power spectrum, which in our case includes the PBH Poisson contribution.  The results are shown in Fig. \ref{fig:bias} (lines) where the effect of the Poisson contribution for high PBH masses and $M^*$ parameters is very strong, as  models with  $M_{\rm PBH}$ or $M^*\sim 10^4$h$^{-1}M_\odot$ (monochromatic, top, and extended PBH mass distributions, middle and bottom, respectively) or greater are clearly inconsistent with observations as haloes of the mass of galaxy clusters would cluster even less strongly than the smooth matter field.  For comparison we show the measurements by \cite{Wang2008} who measure the clustering of groups of galaxies; in particular we show the results for masses calibrated using the abundance matching technique, i.e. using numerical simulations to match the abundance of dark matter haloes to that of galaxy groups, ranking the former by their dark matter mass and the latter by their total luminosities.  Even though the masses are somewhat dependent on the fiducial cosmology, the increase in the level of clustering with halo mass is clear.  \cite{Wang2008} present bias as a function of group mass relative to their lowest mass sample; because of this we rescale their results to the amplitude of $\Lambda$CDM at their lowest mass bin.

The drop of the clustering amplitude  to values even below that of the mass ($b<1$) for high values of PBH mass or $M^*$ in the PBH mass function shows that  haloes are forming primarily from Poisson fluctuations sourced by massive PBHs (individual ones for monochromatic functions); i.e. the power spectrum at the scales of collapse of massive haloes contains a significant contribution from Poisson fluctuations (cf. Figures \ref{fig:pkpmon} and \ref{fig:pk}) which are spatially uncorrelated and therefore tend to wash out the $\Lambda$CDM clustering, in stark contrast with the  trend shown by galaxy groups.  The excess of power on scales corresponding to the mass of these large haloes makes their collapse much more common than in $\Lambda$CDM.  In the latter, such objects would be located on extreme peaks of the density field and cluster more strongly (see for instance \citealt{Padilla2002,Wang2008,Finoguenov2020}). 

When the PBH mass, or the characteristic mass drops to $10^3$h$^{-1}M_\odot$, the strong effect of Poisson fluctuations vanishes and the $\Lambda$CDM model behaviour is recovered, modulo $10\%$ differences below $M_{h}=10^{13}$h$^{-1}M_\odot$.  In the following section we will quantify the departure in the clustering dependence on mass from observations with respect to $\Lambda$CDM, to assess the region of parameter space that produces halo clustering that matches observations as reasonably well as is achieved with the standard cosmological model.

\begin{figure}
    \centering
    \includegraphics[width=0.4\textwidth]{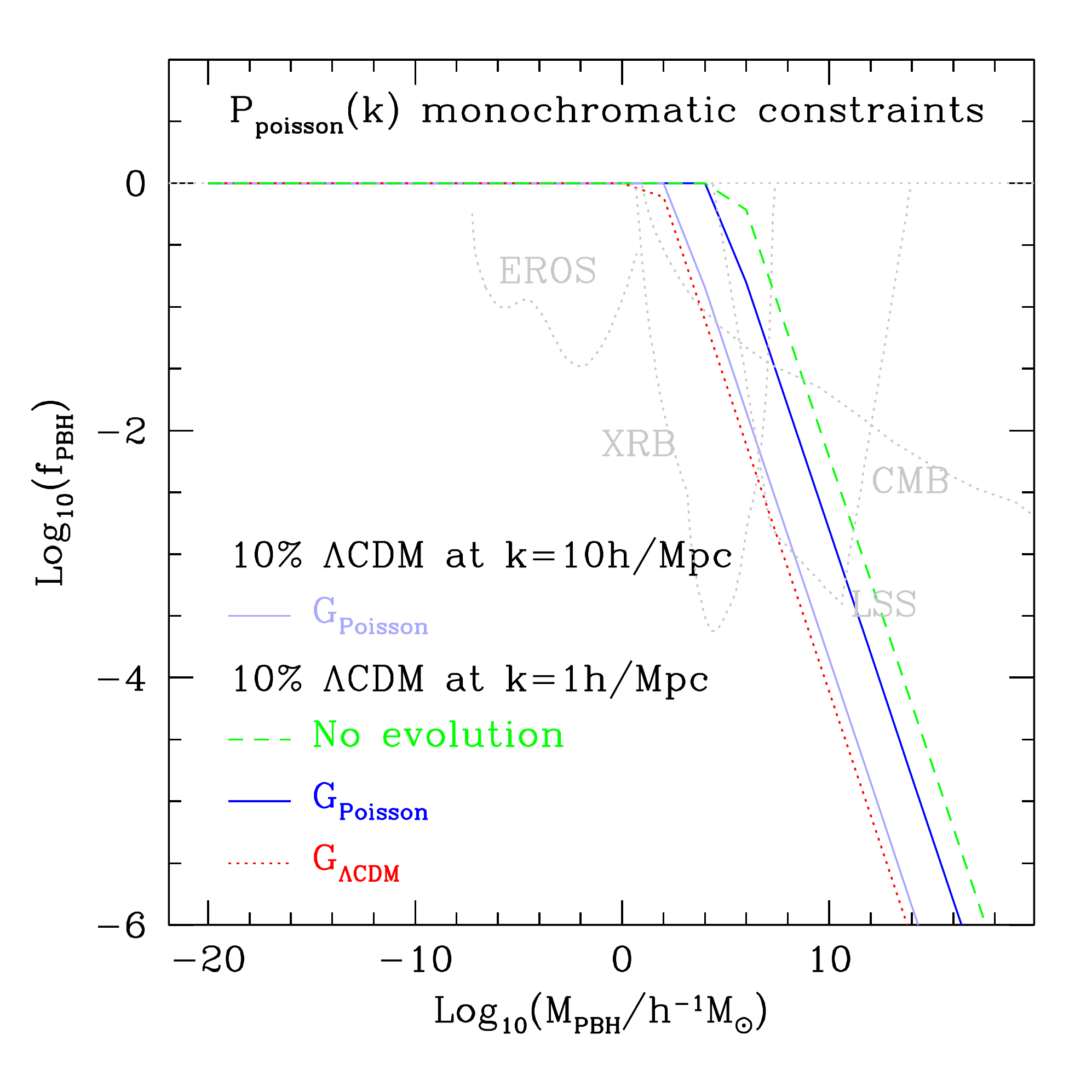}
    \includegraphics[width=0.4\textwidth]{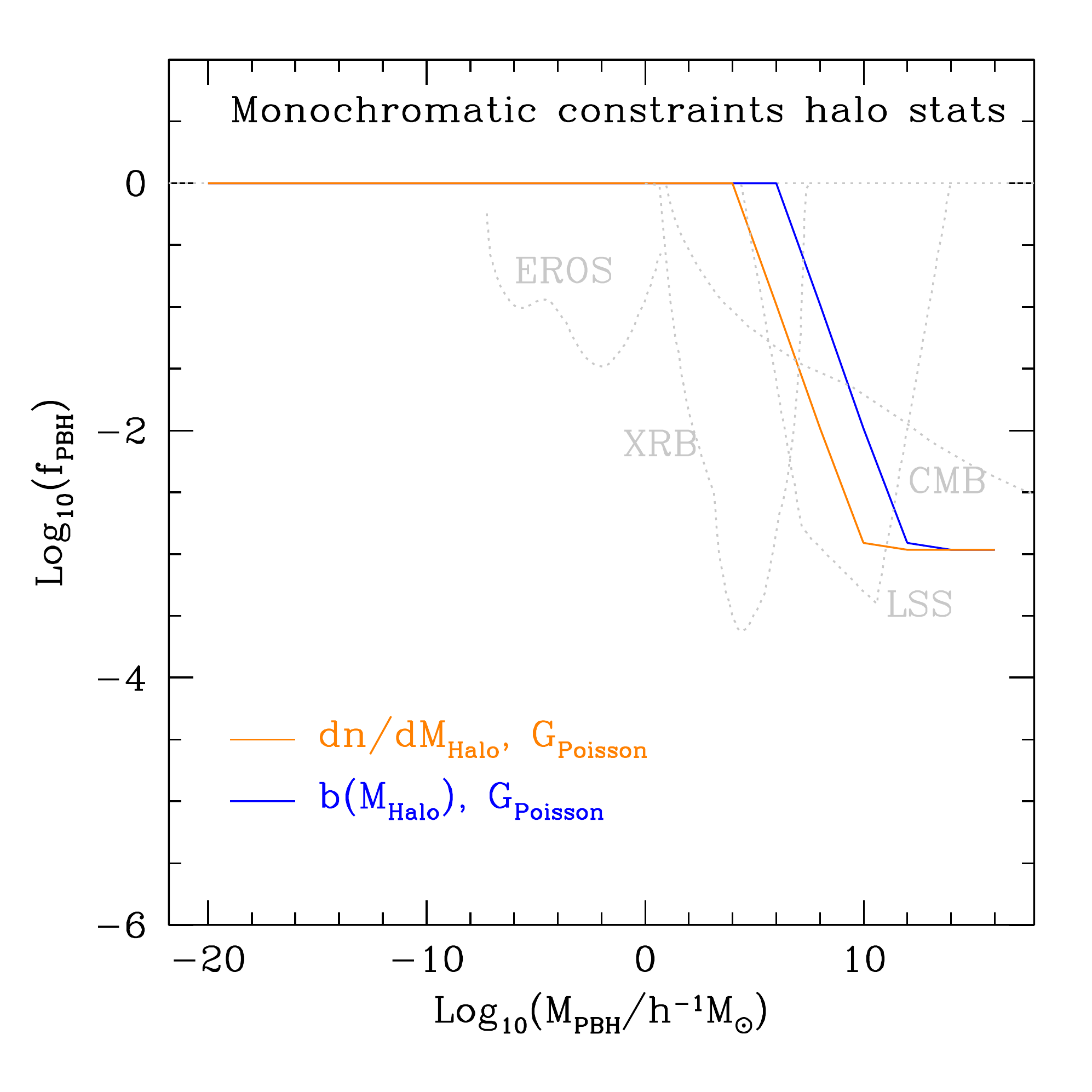}
    \vskip-.3cm
    \caption{Fraction of dark matter in PBHs as a function of monochromatic PBH mass, from the contribution of Poisson noise to the linear power spectrum (top) and to the halo mass function and bias as a function of mass (bottom). In the top panel, the solid line shows the case of growth since the redshift of matter and radiation equality, whereas the dashed is the case of no growth and dotted, of growth since the mode enters the horizon. The light blue solid line shows the more stringent case of requiring a small variation of the power spectrum with respect to $\Lambda$CDM at $k\leq 10$hMpc$^{-1}$. In the bottom panels the blue line shows the constraints from the halo bias as a function of mass, whereas the orange line shows the constraints from variations in the halo mass function; both cases correspond only to growth during matter domination for clarity.  The grey dotted lines in both panels show a selection of other constraints for monochromatic PBH mass distributions from lensing (EROS, \citealt{Tisserand:2007}), x-ray binaries (XRB, \citealt{Inoue_2017}), effects of accretion on the CMB \citep{Serpico_2020}, and large scale structure effects (LSS, extracted from \citealt{Carr_2018}).}
    \label{fig:fpkpmon}
\end{figure}

\section{Constraints on the abundance of PBH and their possible role as dark matter}
\label{pofk}

{Here we provide quantitative constraints on the possibility that dark matter is in the form of PBHs in either monochromatic or PS mass functions.

As there is no evidence of departure from $\Lambda$CDM in the observed CMB temperature power spectrum \citep{PlanckX2018}, any PBH population is required to satisfy,
\begin{equation}
    f^2_{\rm PBH}P_{\rm Poisson}^{\rm PBH}(k,z)<<P_{\Lambda {\rm CDM}}(k,z), \text{for } k\leq k_{\rm NL}.
\label{eq:poissonmono}
\end{equation}
The upper limit on $k$ for this condition corresponds to the scale where non-linear, baryonic and redshift space distortion effects make it difficult to predict the actual measured $P(k)$ by the model.  We choose $k_{\rm NL}=1$ and $10$hMpc$^{-1}$ (in some cases).  The first scale is where the effect of baryons is generally thought to reduce the non-linear power spectrum amplitude by about $30\%$ \citep{Schneider_2019}. The larger $k_{\rm NL}$ value corresponds to the scale at which a warm dark matter model starts to depart from CDM \citep{Stuecker:2018}.  Apart from baryons and the nature of DM, primordial non-gaussianity can also affect the linear power spectrum on small scales.  \cite{Taruya_2008} show that the effect at these scales is of a few percent at low redshifts.  The main effect of non-linear evolution is to broaden much more strongly the positive tail of the distribution of overdensities thereby increasing the amplitude of the power spectrum by orders of magnitude, making it difficult to find traces of small variations of the power spectrum at high wavenumbers.
However, in the last equation, we will still adopt a conservative maximum of $10\%$ effect on the linear $P_{\Lambda {\rm CDM}}(k)$, but will make the comparison between the Poisson and $\Lambda$CDM power spectra at $z=1100$ when fluctuations are still linear.

Substituting and solving for $f_{\rm PBH}$ at $k_{\rm NL}$,
\begin{equation}
    f_{\rm PBH}(M,z)=\min\left(1,\left(\frac{0.1 \ P_{\Lambda {\rm CDM}}(k_{\rm NL},z)}{P_{\rm Poisson}^{\rm PBH}(k,z) }   \right)^{1/2}\right),
    \label{eq:fpkmon}
\end{equation}
where the $0.1$ factor corresponds to the adopted $10\%$ effect.

\begin{figure*}
    \centering
    \includegraphics[width=0.45\textwidth]{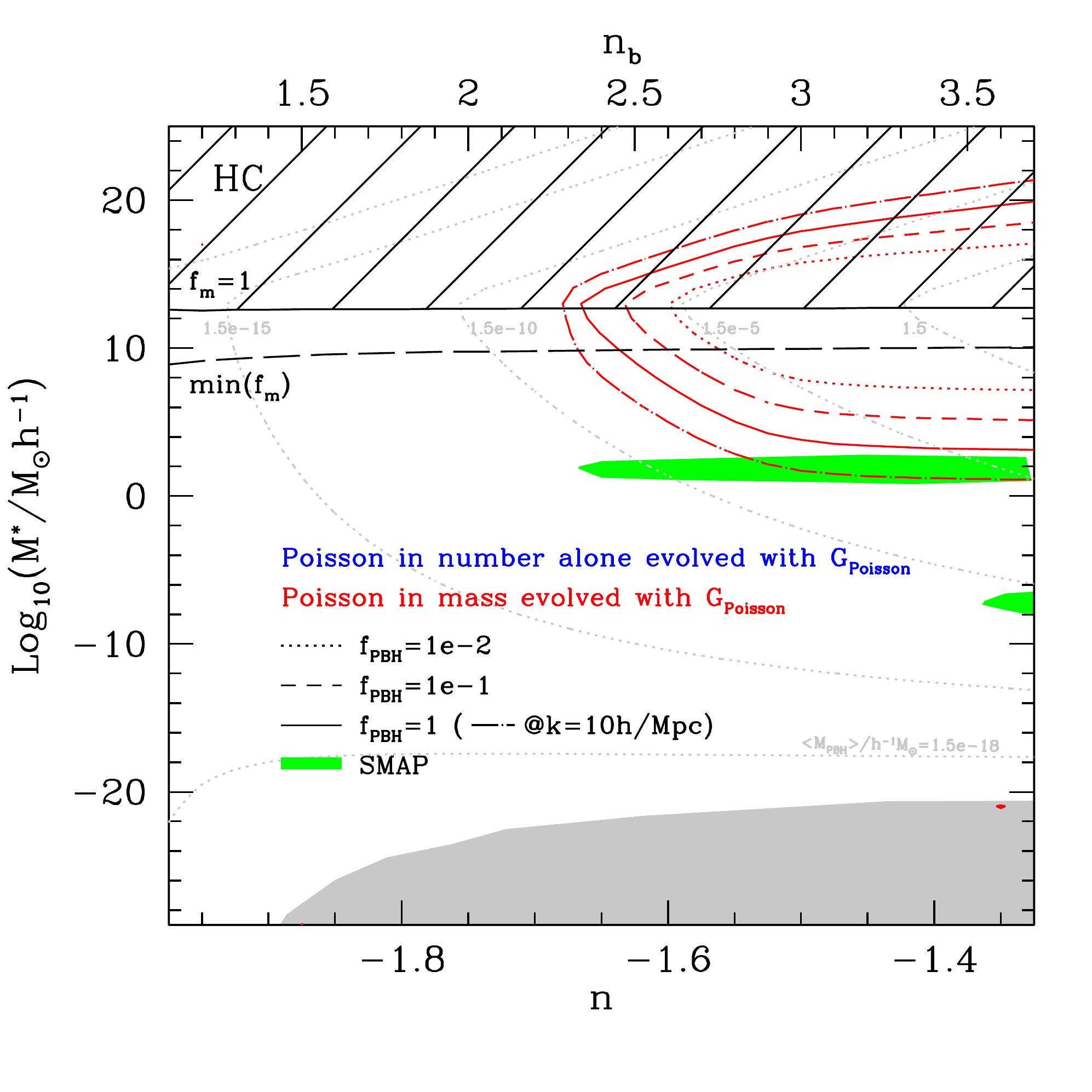}
    \hskip -1.6cm \includegraphics[width=0.45\textwidth]{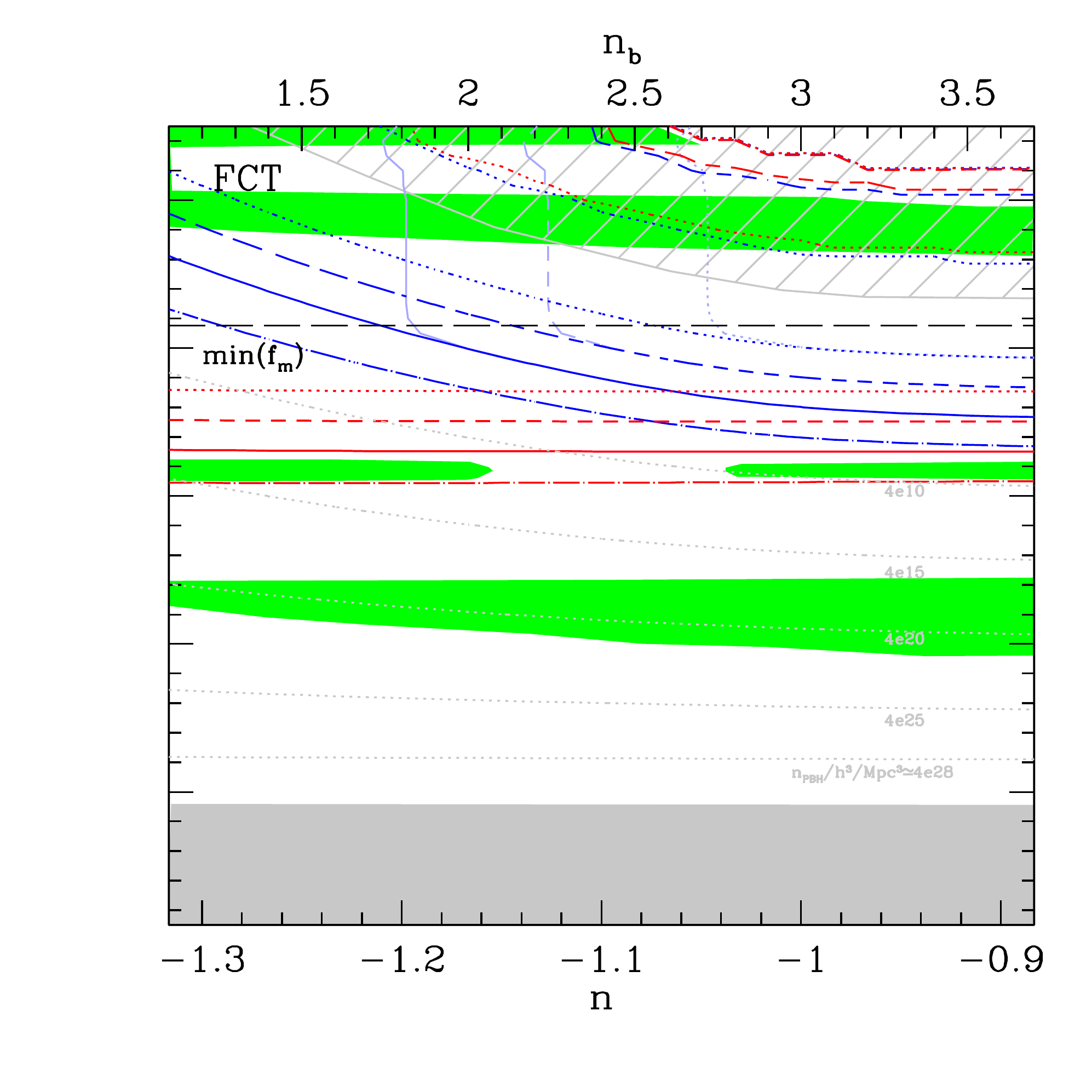}
    \vskip-.5cm
    \includegraphics[width=0.45\textwidth]{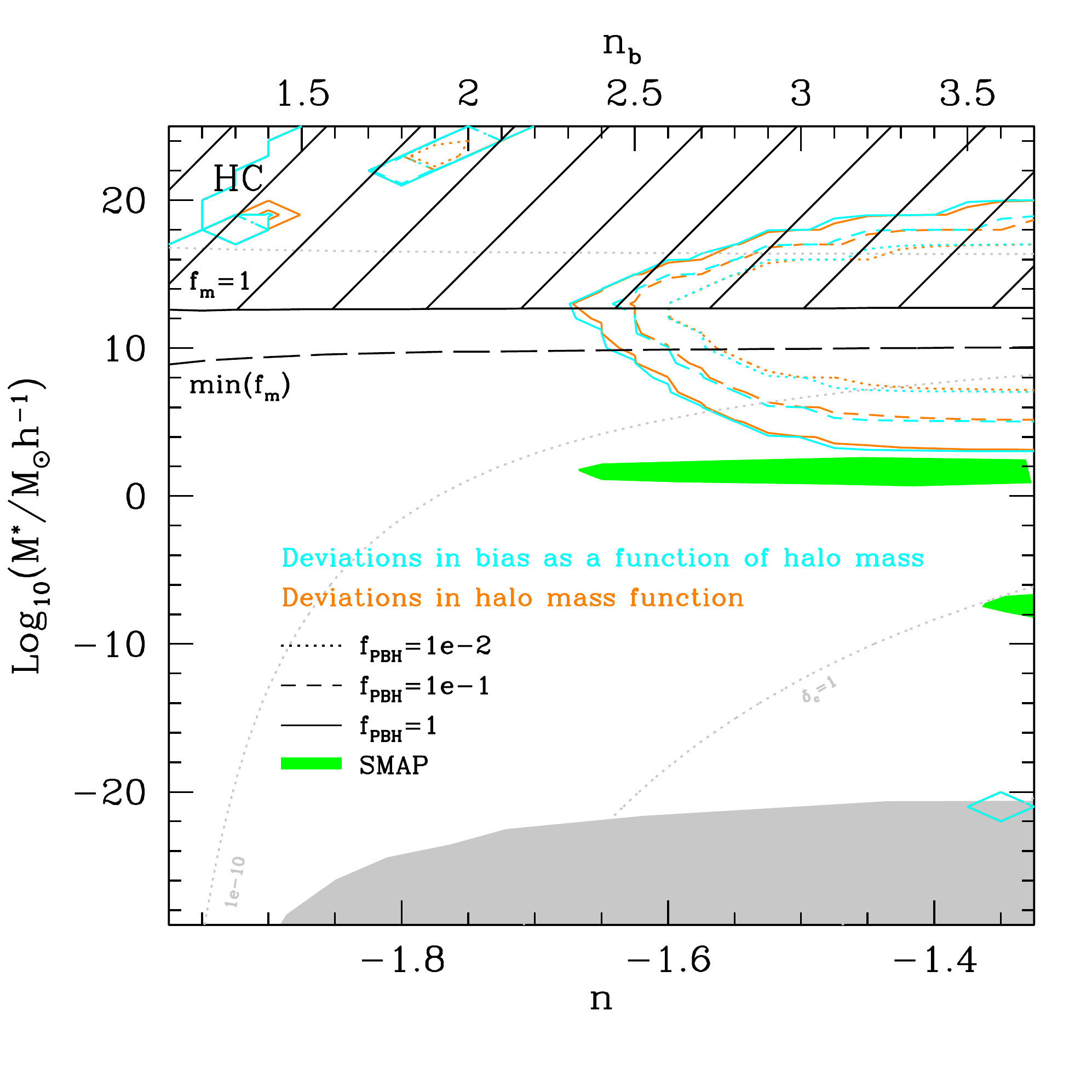}
    \hskip -1.6cm \includegraphics[width=0.45\textwidth]{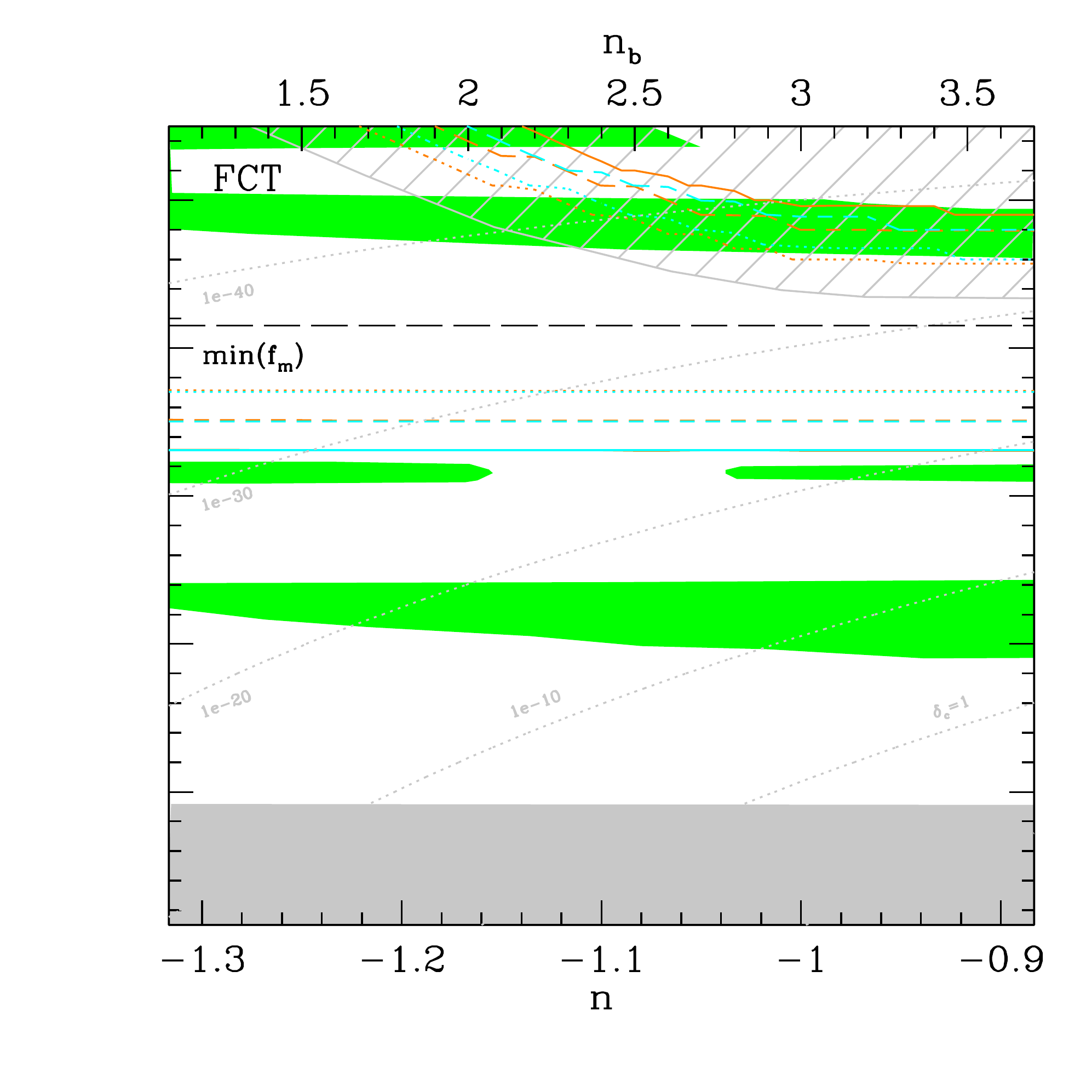}
    \vskip-.5cm
    \caption{Top: Fraction of dark matter in PBHs as a function of the PBH mass function parameters $n$ and $M^*$, resulting from Eq. \ref{eq:fpkmon} applied to the extended mass function poisson power spectra of Eq. \ref{Eq:pk}, taking only into consideration the case of growth of poisson fluctuations during matter domination. The different line types show different fractions of matter in PBH (see the key). Red colours show the poisson noise on both, number density and matter, whereas blue corresponds to  poisson noise from number densities alone. The left and right panels show HC and FCT mass functions for $f_m=1$, respectively, for the same range of blue spectral indices (shown on the top axis).  The powder blue lines on the top-right correspond to FCT with $\min(f_m)$ (shown only for Poisson in number as there is no difference for Poisson in mass between the two FCT cases).   All lines correspond to fractions calculated at $k_{\rm NL}=1$hMpc$^{-1}$, except the long dashed-dotted lines, which show the $f_{\rm PBH}=1$ contour for $k_{\rm NL}=10$hMpc$^{-1}$.  The grey contours show different values of the average PBH mass (left) and PBH number density (right) as labelled.  Bottom: Fraction of dark matter in PBHs such that the reduced $\Delta\chi^2=1$ between the halo mass function calculated using the full poisson power spectrum and the $\Lambda$CDM halo mass function with observational errors from \protect\cite{Abdullah2020} at equal abundance (orange), and the same for the bias as a function of halo mass compared to \protect\cite{Wang2008} (cyan), as indicated in the legend.  The grey contours in the bottom panels show the values of linear overdensity for collapse, as labelled.  In the top and bottom panels the  near horizontal black line in the left panels shows the pivot mass scale $M_{\rm pivot}$ above which there are no possible mass functions for HC (hatched region); the long dashed black line shows the case of the minimum possible $\min(f_m)$, which further limits $M^*$ to take this maximum value in HC (left), and marks the pivot feature in the mass function for the $\min(f_m)$ FCT case (right).  The grey hatched regions in the right panels correspond to departures from $\Lambda$CDM due to lack of fluctuations on $k_{\rm NL}=1$hMpc$^{-1}$. The green shaded regions show the parameter ranges that allow $100\%$ of mass in PBH according to \protect\cite{Sureda:2020}.  The grey area shows the region where there are no PBH at $z=0$.}
    \label{fig:fpk}
\end{figure*}

To calculate the maximum allowed fraction of dark matter in PBHs from changes in the halo mass function and halo bias as a function of mass we use the standard $\chi^2$ per degree of freedom (d.o.f.) assuming no correlation between mass bins,
$$\chi^2(\Theta)=\frac{1}{n_{d.o.f.}}\sum_i \left(\frac{(O(M_i)-{\cal M}(\Theta,M_i)}{\sigma_O(M_i)}\right)^2,$$
where $O(M_i)$ and $\sigma_O(M_i)$ correspond to the measurements and errors of the halo bias or halo mass function in bins of halo mass $M_i$, and ${\cal M}$ is the corresponding model that depends on the mass function parameter set $\Theta$, which is $\Theta=\{M_{\rm PBH}\}$ for the monochromatic case, and $\Theta=\{M^*,n_b\}$ for the extended PS distributions.  The maximum allowed fraction of mass in PBHs is then the one that produces a departure from $\{O(M_i)\}$ of $\Delta \chi^2=\chi^2-\chi^2_{\rm min}=1$ with respect to the minimum reduced $\chi^2$ value.

For the bias as a function of mass we only use the three highest mass bins of \cite{Wang2008} as $\{O(M_i)\}$ because the first mass bin is only used as a reference amplitude and is quoted with zero error.

For the measurements of the halo mass function from \cite{Abdullah2020} we notice that the halo mass measurements in the lower mass bins depart significantly from the $\Lambda$CDM case (cf. Fig. \ref{fig:hmf}) which induces a very bad fit for models with and without PBHs.  Since we do not intend to allow $\Lambda$CDM parameters to vary in this work, instead of comparing the halo mass functions of models and observations, we take the errors of the measurements and apply them to the $\Lambda$CDM halo mass function at equal number density.  We take this as the $\{O(M_i)\}$ set of data and repeat the procedure followed for the bias of haloes as a function of mass. }

\subsection{Constraints on $f_{\rm PBH}$ for monochromatic distributions}

The effect of Poisson fluctuations coming from single mass black holes was presented in Fig. \ref{fig:pkpmon} for the linear power spectrum and in the top panels of Figs. \ref{fig:hmf} and \ref{fig:bias} for the halo mass function and dependence of halo bias on halo mass, where in all cases the effect becomes larger as the PBH mass increases.  

Applying the criterion that the modification to the $\Lambda$CDM power spectrum at wavenumbers of $1$ and $10$hMpc$^{-1}$ should not exceed $10\%$ returns allowed fractions of DM in PBH  from Eq. \ref{eq:fpkmon} which are shown in the top panel of Fig. \ref{fig:fpkpmon}.  As can be seen, for $k=1$hMpc$^{-1}$ the least permissive case is that of growth since the mode enters the horizon, as expected, although this is only a shift of $2$dex on the PBH mass that allows $100\%$ of DM in PBHs (ignoring any other constraints) with respect to the preferred case of growth since matter and radiation equality.  The upper limit for $f_{\rm PBH}=1$  ranges from $M_{\rm PBH}=10^2$h$^{-1}M_\odot$  to $M_{\rm PBH}\sim 10^5$h$^{-1}M_\odot$ in the case of no evolution.  We also show the constraints with growth since equality ($G_{\rm Poisson}$) for $k=10$hMpc$^{-1}$ as a light blue line, and restrict masses to $M_{\rm PBH}<10^2$h$^{-1}M_\odot$ in order to allow $f_{\rm PBH}=1$.  Incidentally, \cite{Gow2020} show that  even with very narrow peaks in the power spectrum it is rather difficult to form PBHs with monochromatic (or narrow log-normal) distributions beyond these masses.  {Notice that at the masses where the approximation adopted for $P(k)_{\rm Poisson}^{\rm PBH}$ for the monochromatic case is no longer accurate ($M_{\rm PBH}\sim 10^{13}$h$^{-1}M_\odot$) the allowed fraction of DM in PBHs is already very low, $<10^{-3}$.}

{
The resulting maximum allowed fractions of dark matter in PBHs resulting from the $\Delta \chi^2$ between the measured dependence of bias on halo mass and the model with different fractions of DM as PBHs are shown as a blue solid line in the bottom panel of Fig. \ref{fig:fpkpmon}.   
The orange line shows the  allowed fractions $f_{\rm PBH}$ for the constraint using the halo mass function.  Both constraints are shown only for growth of Poisson fluctuations since matter domination. }

In both the top and bottom panels, we further compare our constraints with a representative set of other monochromatic constraints  for the same region of $M_{\rm PBH}$ (there are other constraints for lower masses that are not shown for brevity), namely, lensing events toward the Magellanic clouds (EROS, \citealt{Tisserand:2007}), from x-ray binaries (XRB, \citealt{Inoue_2017}), effects coming from the accretion of PBHs and how this affects the CMB \citep{Serpico_2020}, and large scale structure effects also from Poisson fluctuations (LSS, extracted from \citealt{Carr_2018}); as it can be seen, the latter place constraints consistent with ours for $G_{\rm Poisson}$ at $k_{\rm NL}=1$h$/$Mpc, as expected, but our constraint from departures of the power spectrum does not taper off for higher masses.  The constraints from the halo mass function and bias as a function of mass for growth since equality (bottom panel) are also consistent with our results for departures from the power spectrum at $k_{\rm NL}=1$h$/$Mpc, and with the LSS constraints even showing a similar minimum allowed fraction of $f_{\rm PBH}\sim 10^{-3}$ for $M_{\rm PBH}>10^{10}$.  In all cases, from our measurements we find strong drops in the allowed fraction of mass in monochromatic PBHs starting at $M_{\rm PBH}=10^4$ to $10^6$h$^{-1}M_\odot$.

\subsection{Constraints on $f_{\rm PBH}$ for extended mass distributions}

In this subsection we look at constrains on the fraction of DM in PBHs with extended mass distributions coming from the effect of Poisson noise on the power spectrum and from departures of the halo mass function and clustering from their observed values.

As presented in Eq. \ref{eq:fpkmon}, we allow the power spectrum to depart by at most $10\%$ in two wavenumbers $k=1$ and $10$hMpc$^{-1}$.  The resulting constraints for the case where Poisson fluctuations can grow during matter domination alone can be seen in the top panels of Figure \ref{fig:fpk}. The case where Poisson fluctuations take into account the change in the mass density due to the extended nature of the mass function is shown in red; this is the case that is more relevant for the extended mass distribution, where apart from noise on the number density, there is also an associated noise of the mass due to high mass PBHs entering or not entering the volume (i.e. the mode $k$).  As can be seen in this case, when requiring a maximum departure of $10\%$ from $\Lambda$CDM at $k=1$hMpc$^{-1}$ the  region of parameter space that allows $100\%$ of dark matter in PBHs corresponds to $M^*<100$h$^{-1}M_\odot$ in both FCT and HC, regardless of the choice of $f_m$, with a less stringent constraint in HC for $n<-1.6$ or, equivalently, $n_b<2.5$.    The constraints ignoring Poisson noise in the mass (blue lines) are non-existent in HC indicating no constrains on the range of parameters shown in the figure.  

Only in FCT the Poisson noise in number alone produces constraints on the parameter space, with $f_{\rm PBH}<1$ regions in the right panel at higher $M^*$ values than for Poisson noise including mass (blue vs. red contours, respectively); this is not unexpected since Fig. \ref{fig:pk} showed that the Poisson noise from number density is lower than when including noise from PBH masses for FCT, and even more so for HC.  The case of FCT with $\min(f_m)$ is even less stringent for the spectrum from Poisson noise in number alone and allows all mass functions with $n_b$ below the intersection of the dark blue fractions with the pivot mass for $\min(f_m)$ (light blue).  The bump in the light blue contours at $M^*\sim 10^{24}$h$^{-1}M\odot$ is due to the tail of PBHs with masses above $M_{\rm piv}$ taking momentary importance before their abundances drop below 1 per horizon as $M^*$ continues to increase (see example mass functions in Fig. \ref{fig:cmf}).  The long dashed-dotted lines show the $f_{\rm PBH}=1$ contour when the restriction is applied at $k=10$hMpc$^{-1}$, and as it can be seen the effect is that of shifting down the contour by about $3$dex in $M^*$ for HC and FCT.

The region with $f_{\rm PBH}\sim 0.1$ allowed in the high $M^*$-$n_b$ range for $f_m=1$ in the upper right panel is rejected for the following reason.  The Poisson power spectrum for PBH mass function parameters in the grey hatched region vanishes for wavenumbers  $<k_{\rm NL}=1$hMpc$^{-1}$ because the PBHs are typically more massive than the horizon mass that corresponds to this scale.  In this case the amount of dark matter on the $\rm NL$ scale as it enters the horizon is different than assumed in $\Lambda$CDM for any $f_{\rm PBH}>0$ and this modifies the total mass power spectrum, which is not accurately represented in our formalism.  

Notice that these constraints rule out the $f_{\rm PBH}=1$ regions (green) with high $M^*$ for FCT found by SMAP.  It also introduces further partial constraints than those found in SMAP for HC for the case of the Poisson power spectrum constraint using $k_{\rm NL}=10$hMpc$^{-1}$, but without completely removing their allowed $f_{\rm PBH}=1$ region with $M^*\sim 100$h$^{-1}M_\odot$.

For illustrative purposes the grey dashed contours in the top panels show the average mass of PBHs, which can in turn be converted into space densities (labels on the left and right show these quantities, respectively).  In the bottom panels, the grey contours show the linear overdensity for collapse.

As Poisson fluctuations can influence the gravitational potential slightly prior to matter domination, we also looked at the more restrictive case of when allowing Poisson fluctuations to grow since entering the horizon, but the changes are small, with shifts in the $f_{\rm PBH}=1$ contours for $k_{\rm NL}=1$hMpc$^{-1}$ of only a fraction of a dex toward lower values of $M^*$, still unable to remove the SMAP allowed ranges for $M
^*<100$h$^{-1}M_\odot$.

The bottom panels of Fig. \ref{fig:fpk} show contours of the fraction of dark matter in PBHs constrained using the halo mass function (orange) and the bias of haloes as a function of their mass (cyan). { In order to estimate the maximum allowed PBH fraction for the bias of haloes we again use the standard $\chi^2$ per degree of freedom between models and observations with $\Theta=\{M^*,n_b\}$.}
The orange and cyan lines of equal PBH fraction delimit the regions where for the quoted $f_{\rm PBH}$ the reduced $\Delta\chi^2=1$ for the halo mass function and halo bias as a function of mass.  

These contours are calculated only for the case of Poisson fluctuation growth since matter-radiation equality, including Poisson noise both in number density and mass.  As it can be seen, the contours show that the constraints from variations in the power spectrum, adopting the condition of $10\%$ departure from the $\Lambda$CDM power spectrum at $k_{\rm NL}=1$hMpc$^{-1}$, are equivalent to the constraints from the halo mass function and halo bias, as the solid contours of the top and bottom panels are an excellent match.

In the case of the mass functions and halo bias, the high $M^*$ FCT regions allowed to conform $100\%$ of DM according to SMAP are also mostly excluded except for the region with $M^*=10^{25}$h$^{-1}M_\odot$ and $n_b\sim 2.5$, but this region is rejected 
as explained above. 

\section{Conclusions}
\label{sec:conclusions}

In this work we studied different aspects of the phenomenology of dark matter clustering when it is composed at least in part by primordial black holes, both in the case where these have a monochromatic mass distribution, i.e. have a single unique mass, and for Press-Schechter-like mass functions in two formation scenarios; (i) collapse of fluctuations as they enter the horizon (Horizon Crossing, HC), and (ii) formation at fixed, early conformal time (FCT).  These cases were presented in \cite{Sureda:2020}.

We first notice that for extended mass functions, there is a change of the mass density in black holes as the volume increases, which allows PBHs of increasingly higher mass to achieve a number density of $1$ per  volume, and therefore increase the mass density as a function of scale.  Another factor that changes the PBH mass density is the evaporation by Hawking radiation, which we assume all black holes are subject to.  This variation of the number density of PBHs per unit volume, as a function of the volume, introduces novel features in the Poisson noise of PBHs that depends on scale in a non trivial way.

We calculate the effect of the Poissonian nature of PBHs, which are tens of orders of magnitude more massive than candidate particles for DM, and find that comparing the Poisson power spectrum to the $\Lambda$CDM one, and requiring the contribution of the former to be at most $10\%$ of the latter at a wavenumber of $k_{\rm NL}=1$hMpc$^{-1}$, the parameter space of both monochromatic and extended mass distributions are constrained further, with monochromatic masses restricted to $M_{\rm PBH}<10^3$h$^{-1}M_\odot$ consistent with previous estimates by \cite{Carr_2018}, and $10^2$h$^{-1}M_\odot$ for the more restrictive case using $k=10$hMpc$^{-1}$.  For extended mass functions of either HC or FCT formation mechanism the characteristic mass is limited to $M^*<100$h$^{-1}M_\odot$ ($30$h$^{-1}M_\odot$ for $k=10$hMpc$^{-1}$), which is also the case for the HC mass function for blue spectral indices $n_b>3$.  These constraints remove the windows where all of DM is in PBHs with FCT mass functions with $M^*>10^{15}$h$^{-1}M_\odot$, reported by SMAP.

We confirm that this choice of both the scale and maximum contribution of the Poisson component of the power spectrum is reasonable, as the halo mass function and the clustering of haloes, as measured by their bias factor, show very strong variations with respect to the standard $\Lambda$CDM case, making them inconsistent with observations if the dark matter were to be made entirely of PBHs with $M^*>100$h$^{-1}M_\odot$ for extended black hole mass distributions or $M_{\rm PBH}>10^4$ for monochromatic PBHs.

It is interesting to note that extended PBH mass distributions can still conform $100\%$ of the dark matter when looking at all constraints combined, including other probes of the existence of PBHs.  Our new constraints are not able to rule out the possibility of all of the dark matter to be constituted by PBHs formed at horizon crossing or at fixed conformal time for $M^*=30$h$^{-1}M_\odot$ (for ranges of $n_b$), and for PBHs formed at fixed conformal time at the end of inflation with  $M^*=10^{-7}$h$^{-1}M_\odot$ (see \citealt{Sureda:2020}).  Note that the latter cannot provide the black holes in the mass ranges detected by LIGO \citep{Abbott:2016blz}; a different mechanism for their formation would be required in this case. Whereas the former mass functions would  produce 
$~3\times10^9$ PBHs of $>30M_\odot$ in massive spiral galaxies, which could conceivably find their way into one another to produce the observed gravitational wave signals.

\section*{Acknowledgements}
We thank Nicola Amorisco, Carlton Baugh, Carlos Frenk, Baojiu Li, Jorge Nore\~na, Marco San Mart\'\i n, Dom\'enico Sapone and Jakub Scholtz for helpful discussions.  This project has received funding from the European Union's Horizon 2020 Research and Innovation Programme under the Marie Sk\l{}odowska-Curie grant agreement No 734374.
NP wants to thank the hospitality of the Institute for Advanced Studies at Durham University (UK) and its Fellows programme, during which part of this work was carried out.
NP, JM and JS acknowledge support from CONICYT project Basal AFB-170002. NP and JS were supported by Fondecyt Regular 1191813.
IJA acknowledges funding from ANID through FONDECYT grant No. 3180620.
The calculations presented in this work were performed on the Geryon computer at the Center for Astro-Engineering UC, part of the BASAL PFB-06 and AFB-170002, which received additional funding from QUIMAL 130008 and Fondequip AIC-57 for upgrades.

\bibliographystyle{mnras}
\bibliography{references} 
\bsp
\label{lastpage}
\end{document}